\documentclass[epj]{webofc}
\usepackage[utf8]{inputenc}
\usepackage[varg]{txfonts}   
\usepackage{booktabs}
\usepackage{xcolor}
\usepackage{float}
\usepackage{graphicx, subfigure}
\definecolor{darkred}{rgb}{0.4,0.0,0.0}
\definecolor{darkgreen}{rgb}{0.0,0.4,0.0}
\definecolor{darkblue}{rgb}{0.0,0.0,0.4}
\usepackage[bookmarks,linktocpage,colorlinks,
    linkcolor = darkred,
    urlcolor  = darkblue,
    citecolor = darkgreen]{hyperref}
\usepackage{subfigure}
\wocname{EPJ Web of Conferences}
\woctitle{Lattice2017}
\begin{document}
%
\selectlanguage{english}
\title{%
RG inspired Machine Learning for lattice field theory
}
\author{%
\firstname{Sam} \lastname{Foreman}\inst{1}\and
\firstname{Joel} \lastname{Giedt}\inst{2} \and
\firstname{Yannick}  \lastname{Meurice}\inst{1}\fnsep\thanks{Speaker, \email{yannick-meurice@uiowa.edu} }\and
\firstname{Judah}  \lastname{Unmuth-Yockey}\inst{1,3}\
}

\institute{
Department of Physics and Astronomy, The University of Iowa, Iowa City, IA 52242, USA
\and
Department of Physics, Applied Physics and Astronomy
Rensselaer Polytechnic Institute,\\
 Troy, NY 12180
\and
Department of Physics, Syracuse University, Syracuse, New York 13244, United States}
\abstract{%

Machine learning has been a fast growing field of research in several areas dealing with large datasets. We report recent
attempts  to use Renormalization Group (RG) ideas in the context of machine learning.
We examine coarse graining procedures for perceptron models designed to identify the digits of the MNIST data.
We discuss the correspondence between principal components analysis (PCA) and RG flows across the transition for worm configurations of the 2D Ising model. Preliminary results regarding the logarithmic divergence of the leading PCA eigenvalue were presented at the conference and have been
improved after. More generally, we discuss the relationship between PCA and observables in Monte Carlo simulations  and the possibility of reduction of the number of learning parameters in supervised learning based on RG inspired hierarchical ansatzes.
  }
\maketitle

\def\beq{\begin{equation}}
\def\enq{\end{equation}}

\section{Introduction}\label{intro}
Machine learning has been a fast growing field of research in several areas dealing with large datasets and should be useful 
in the context of Lattice Field Theory \cite{ng}.
In these proceedings, we briefly introduce the concept of machine learning. We then report attempts to use
Renormalization Group (RG) ideas for the identification of handwritten digits \cite{mnist}.
We review the  multiple layer perceptron \cite{rosep} as a simple method to identify digits with a high rate of success (typically 98 percent). We discuss the Principal Components Analysis (PCA) as a method to identify {\it relevant} features. We consider the effects of PCA projections and coarse graining procedures
on the success rate of the perceptron. The identification of the MNIST digits is not really a problem where some critical behavior can be reached. In contrast, the two-dimensional (2D) Ising model near
the critical temperature $T_c$ offers the chance to sample the high temperature contours (``worms" \cite{PhysRevLett.87.160601}) at
various temperature near $T_c$. At the conference, we gave preliminary evidence that the leading PCA eigenvalue of the worm pictures has a logarithmic singularity related in a precise way to the
singularity of the specific heat. We also discussed work in progress relating the coarse graining of the worm images to approximate procedure in the Tensor Renormalization Group (TRG) treatment of the
Ising model \cite{prb87,efratirmp,prd88}. After the conference, much progress has been made in regard to this question. We briefly summarize the content of an upcoming preprint about this question \cite{inprogress}.

\section{What is machine learning?}

\begin{figure}[h]
\begin{center}
    \includegraphics[width=0.6\textwidth]{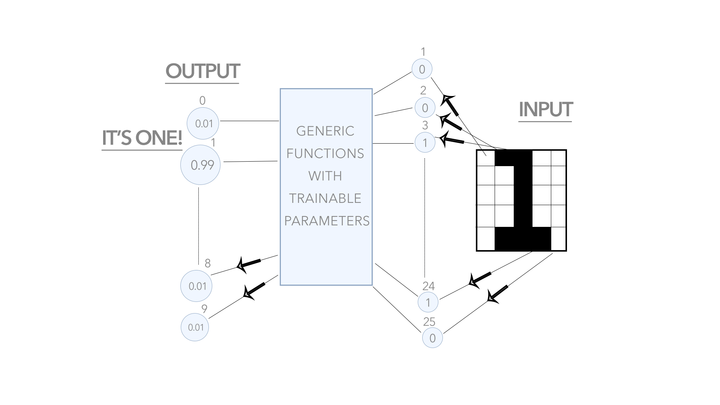}
\end{center}
\caption{\label{fig:rose}Schematic representation of Frank Rosenblatt 's  perceptron \cite{rosep}.}
\end{figure}

If you open the web page of a Machine Learning (ML) course, you are likely to find the following definition:
``machine learning is the science of getting computers to act without being explicitly programmed"
(see e.g. Andrew Ng's course at Stanford \cite{ng}). You are also likely to find the statement that
in the past decade, machine learning has been crucial for self-driving cars, speech recognition, effective web search, and the understanding of the human genome. We hope it will also be the case for Lattice Field Theory. From a pragmatic point of view, ML amounts to constructing functions that provide features (outputs) using data (inputs).
These functions involve ``trainable parameters" which can be determined using a ``learning set" in the
case of supervised learning.
The input-output relation  can be written in the generic form ${\bf y}({\bf v},{\bf W})$ with ${\bf v}$ the inputs, ${\bf y}$ the outputs and ${\bf W}$ the trainable parameters. This is illustrated in Fig. \ref{fig:rose} as a schematic representation of the so-called perceptron \cite{rosep} where the outputs functions have the form $y_l=\sigma(\sum_jW_{lj}v_j)$ with $v_j$ the pixels, $W_{lj}$ the tunable parameters and $\sigma(x)$ the sigmoid function defined below. This simple parametrization allows you to recognize correctly 91 percent of the digits  of the testing set of the MNIST data (see section \ref{sec:mnist}).


\section{The MNIST data}
\label{sec:mnist}
\begin{figure}[hh]
\begin{center}
\vskip-5pt
\includegraphics[width=1.2in]{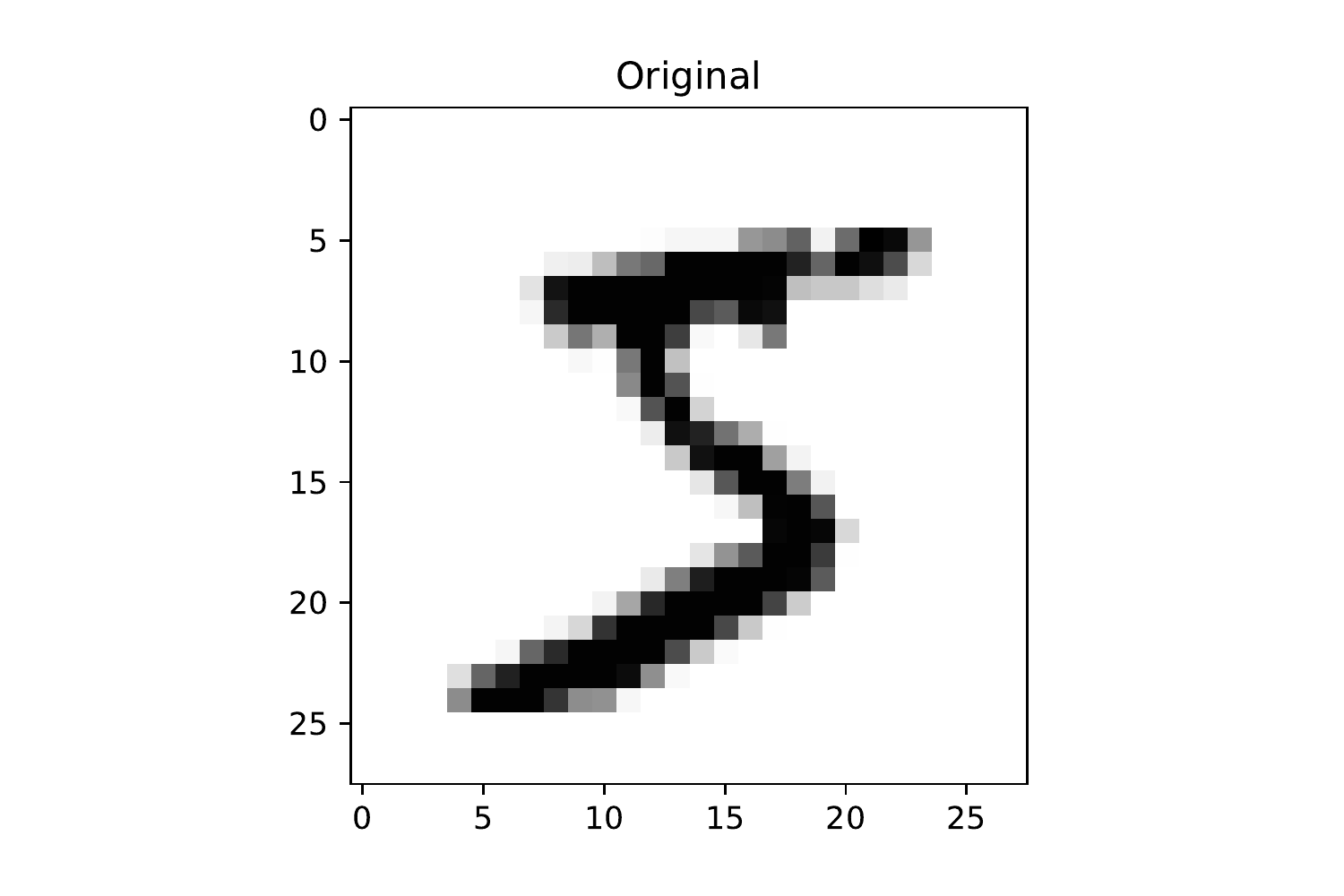}
\includegraphics[width=1.2in]{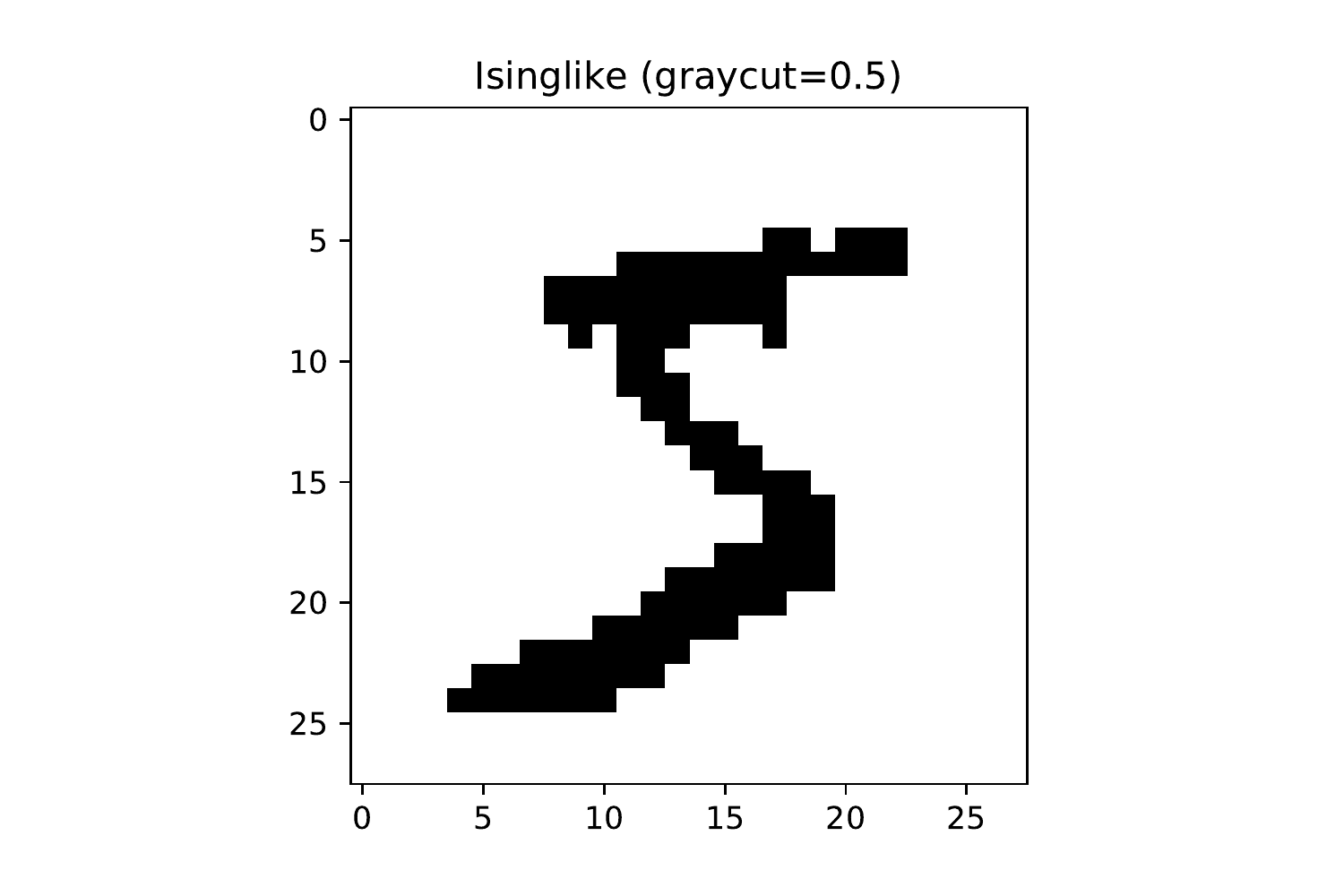}
\includegraphics[width=1.2in]{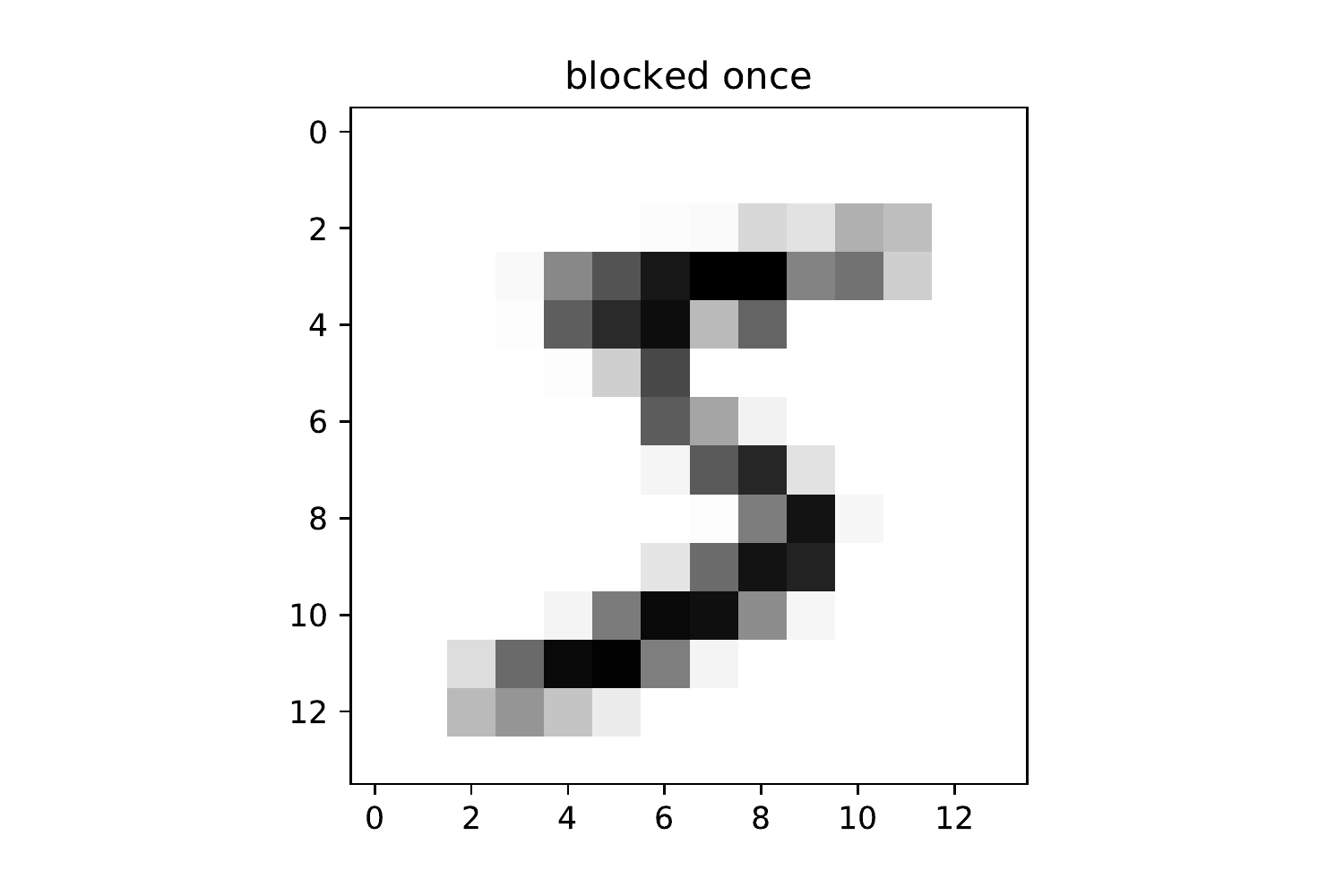}
\includegraphics[width=1.2in]{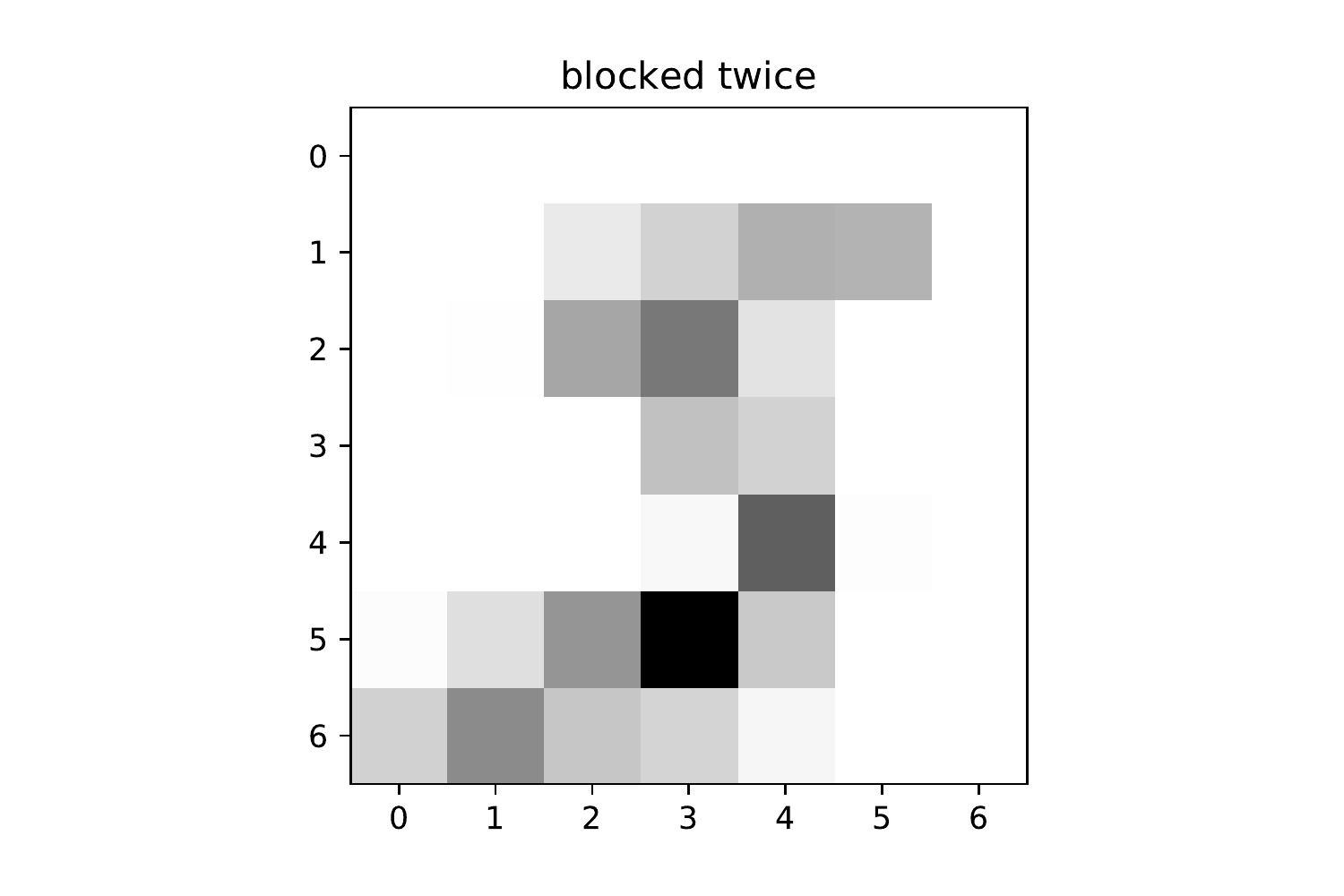}

\end{center}
\label{fig:ising5}
\caption{An example of MNIST image, a black and white approximation with a graycut at 0.5, the image after one and two blockings (left to right). On the rightmost picture, the blocks should appear sharply otherwise you need to use another viewer, Acrobat is showing it properly.}
\end{figure}

A classic problem in ML is the identification of handwritten digits. There exists a standard learning  set called the MNIST data \cite{mnist}. It consists of
60,000 digits where the correct answer is known for each. Each image is a square with $28\times28$ grayscale pixels.
There is a UV cutoff (pixels are uniform), an IR cutoff (the linear size is 28 lattice spacings)  and a typical size (the width of lines is typically 4 or 5). Unless you are attempting to get a success rate better than 98 percent, you may use a black and white approximation and consider the images as Ising configurations: if the pixel has value larger than some gray cutoff (0.5 on Fig. \ref{fig:ising5}), the pixel is black, and white otherwise. Another simplification is to use a blocking, namely replacing groups of four pixels forming a 2 by 2 square by their average grayscale. The blocking process can only be repeated 5 times, after that, we
obtain a uniform grayscale that makes the identification of the digit difficult.

We now consider a simple model, called the perceptron, which generates 10 output variables, one associated with each digit, using functions of the pixels (the visible variables) with one intermediate set of variables called the hidden variables.
The visible variables  $v_i$ are the $28\times28=784$ pixel's grayscale values between 0 and 1. We decided to take 196=784/4 hidden variables $h_k$. Later (see hierarchical approximations in section \ref{sec:rg}), we will ``attach" them rigidly to $2\times2$ blocks of pixels. The hidden variables are defined by a linear mapping followed by  an activation function $\sigma$.
\beq
h_k=\sigma(\sum_{j=1}^{784}W_{kj}^{(1)}v_j), k=1, 2\dots 196.
\label{eq:hlin}
\enq
We choose the sigmoid function  $\sigma(x)= 1/(1+\exp(-x))$, a popular choice of activation function which is 0 at large negative input, 1 at large positive input. We have $\sigma(x)' =\sigma(x)-\sigma(x)^2$  which allows
simple algebraic manipulations for the gradients.
The output variables are defined in a similar way as functions of the hidden variables
\beq
y_l=\sigma(\sum_kW_{lk}^{(2)}h_k)).
\enq
with $l=0, 1, \dots 9$ which we want to associate with the MNIST characters by having  target values $y_l\simeq 1$  for $l=$digit while $y_l\simeq 0$ for the 9 others. The trainable parameters $W_{kj}^{(1)}$ and $W_{kj}^{(2)}$ are determined by gradient search. Given the MNIST learning set $\{ v_i^{(n)}\}$ with $n=1,2, \dots N\simeq 60,000$ with the corresponding  target vectors $\{ t_l^{(n)} \} $, we minimize the loss function:
\beq
{\mathcal E}(W^{(1)},W^{(2)})=(1/2)\sum_{n=1}^N\sum_{l=0}^9(y_l^{(n)}-t_l^{(n)})^2.
\enq
The weights matrices $W^{(1)}$ and $W^{(2)}$ are initialized with random numbers following a normal distribution. They are then optimized using a gradient method with gradients:
\beq
{\mathcal G}^{(2)}_{lk}\equiv\frac{\partial {\mathcal E}}{W_{lk}^{(2)}}=(y_l-t_l)(y_l-y_l^2)h_k
\enq
\beq
{\mathcal G}^{(1)}_{kj}\equiv\frac{\partial {\mathcal E}}{W_{kj}^{(1)}}=\sum_l(y_l-t_l)(y_l-y_l^2) W_{lk}^{(2)}(h_k-h_k^2)v_j
\enq

After one ``learning cycle" (going through the entire MNIST training data), we get a performance of about 0.95 (number of correct identifications/number of attempts on an independent testing set of 10,000 digits),
after 10 learning cycles, the performance saturates near 0.98 (with 196 hidden variables and learning parameters, which control the gradient changes, 0.1 and 0.5.
It is straightforward to introduce more hidden layers:
$y_l^{(n)}\equiv\sigma(W^{(m)}....\sigma(W_{lk}^{(2)}\sigma(W_{kj}^{(1)}v_j^{(n)})).$
With two hidden layers, the performance improves (but only very slightly).
On the other hand, if we remove the hidden layer, the performance goes down to about 0.91 as mentioned above.
It is instructive to look at the outputs for the 2 percent of cases where the algorithm fails to identify the correct digits.
There are often cases where humans would have hesitations.
In the following, we will focus more on getting comparable performance with less learning parameters rather than attempting to reduce
the number of failures.

It has been suggested \cite{schwab14,schwab16} that the hidden variables can be related to the RG ``block variables" and  that an hierarchical organization inspired by physics modeling could drastically reduce the number of learning parameters. This may be called `cheap" learning \cite{teg16}.
The notion of criticality or fixed point has not been identified precisely on the
ML side. It is not clear that the technical meaning of  ``relevant", as used in a precise way in the RG context to describe unstable directions of the RG transformation
linearized {\it near a fixed point}, can be used generically in ML context.
The MNIST data does not seem to have ``critical" features and we will reconsider this question for the more tunable Ising model near the critical temperature (see section \ref{sec:ising}).


\section{Principal Component Analysis (PCA)}
\begin{figure}[h!]
\begin{center}
 \vskip-1cm
\includegraphics[width=4cm]{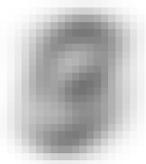}
\vskip -1cm
\caption{\label{fig:ave}Average of each of the 784 pixels over the MNIST learning set.}
\end{center}
\end{figure}

\begin{figure}[b!]
\begin{center}
\includegraphics[width=1.3in]{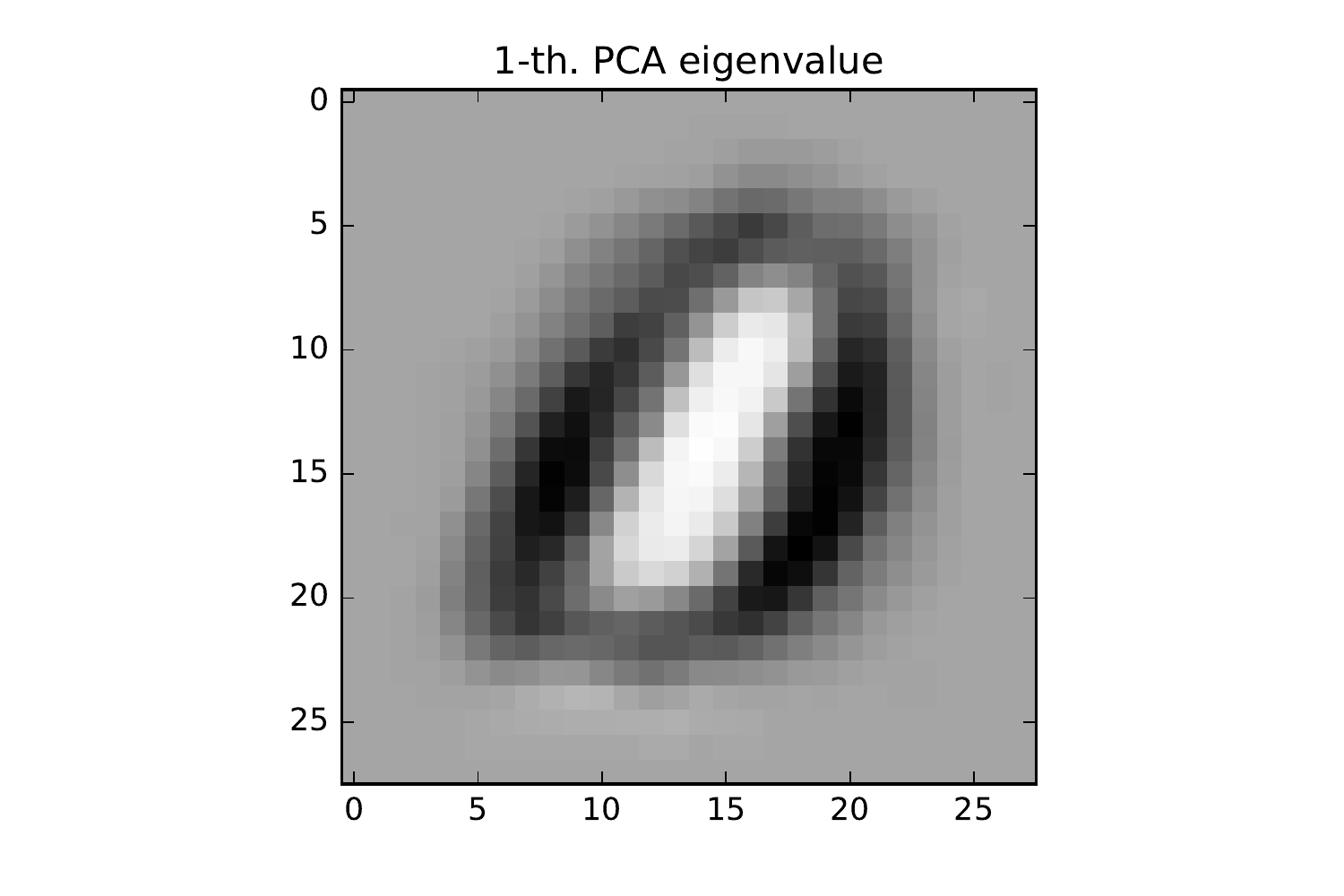}
\includegraphics[width=1.3in]{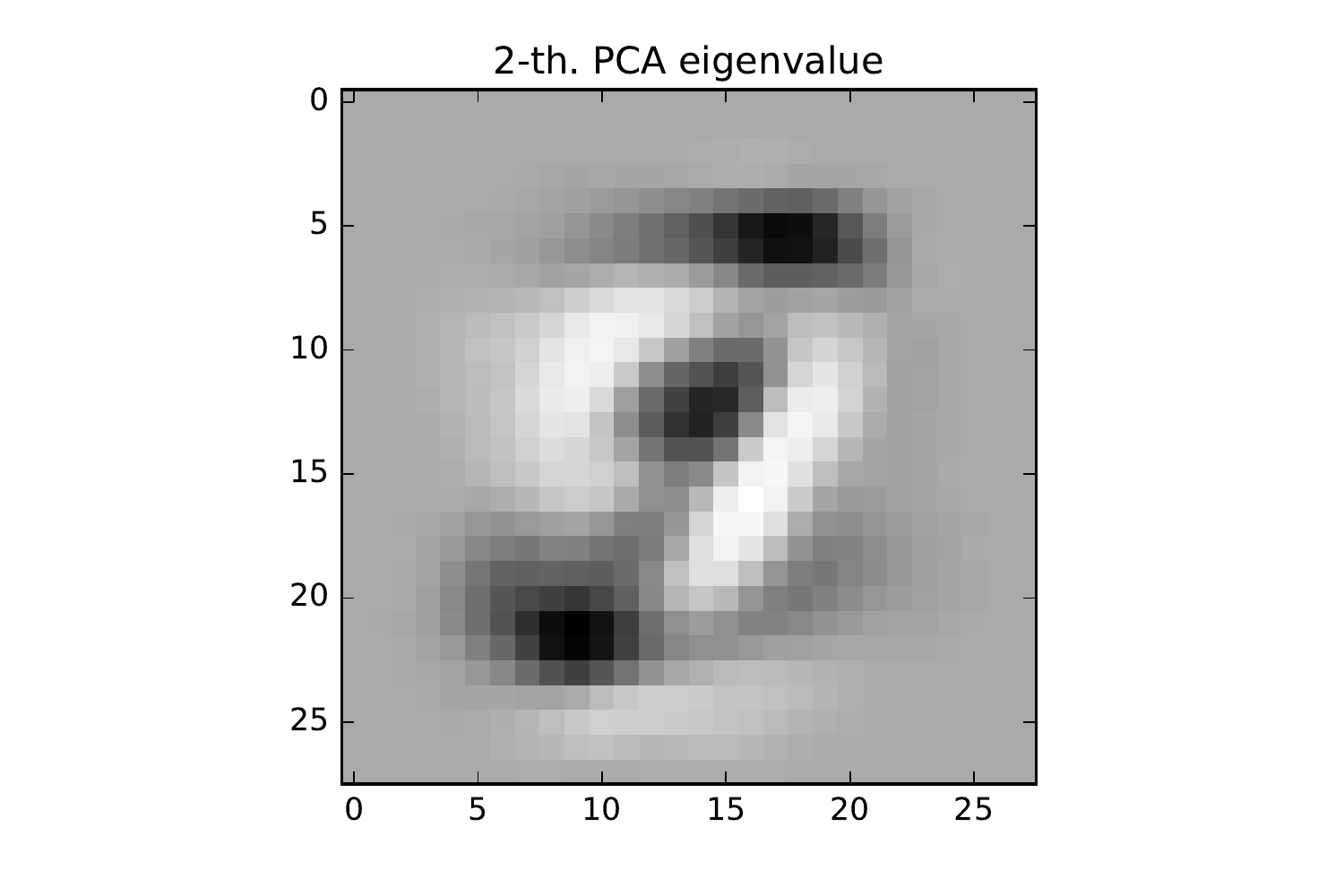}
\includegraphics[width=1.3in]{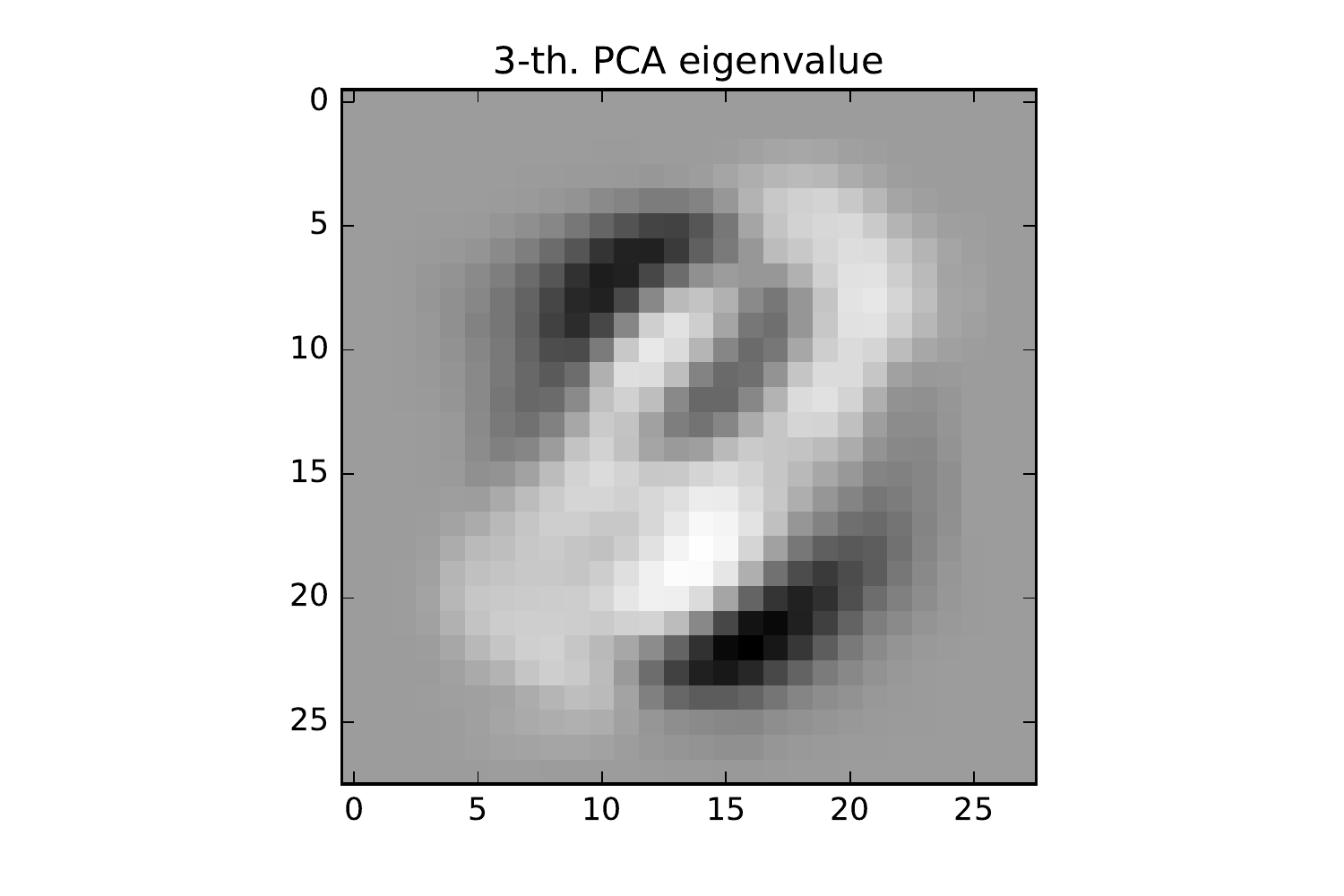}\\
\includegraphics[width=1.3in]{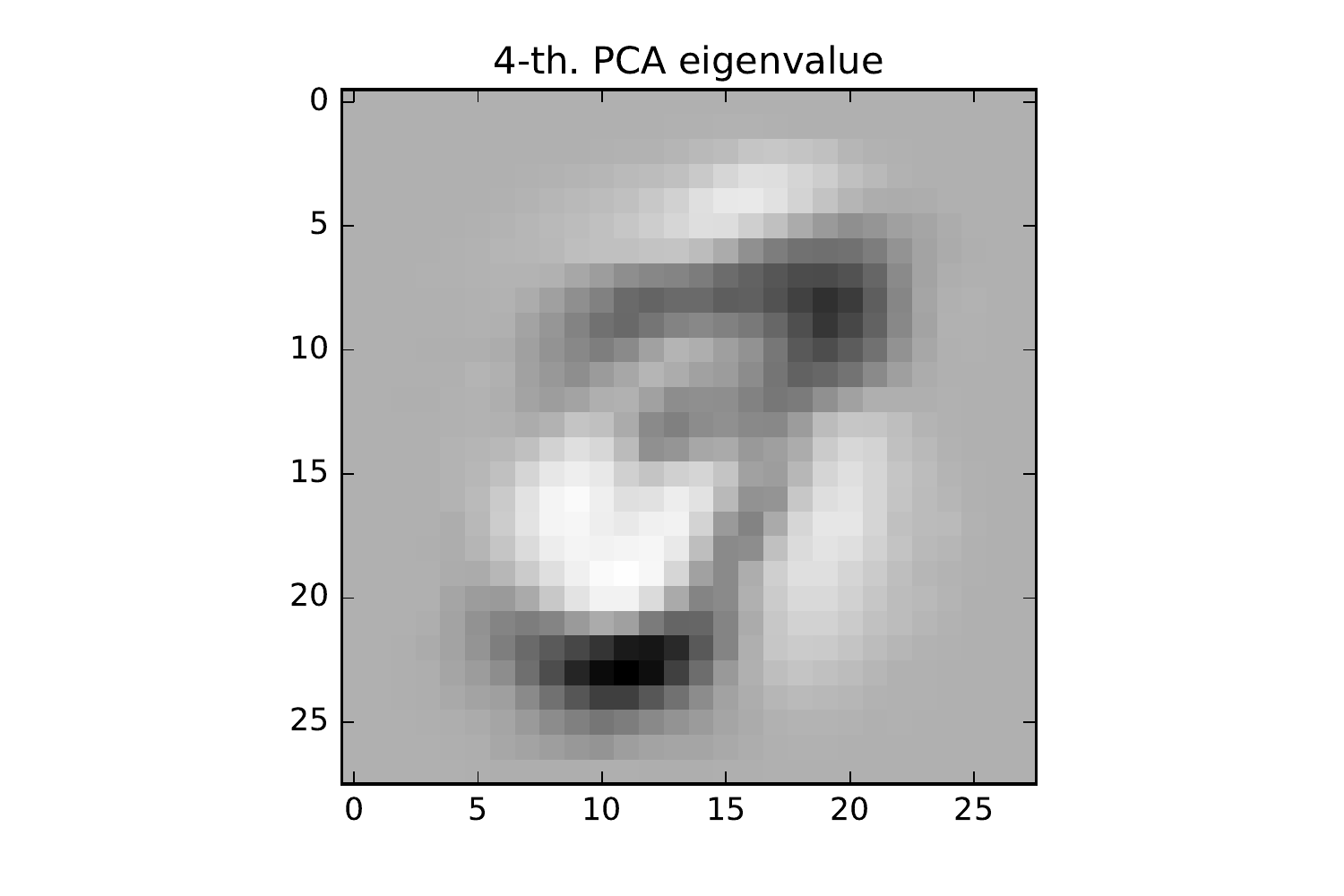}
\includegraphics[width=1.3in]{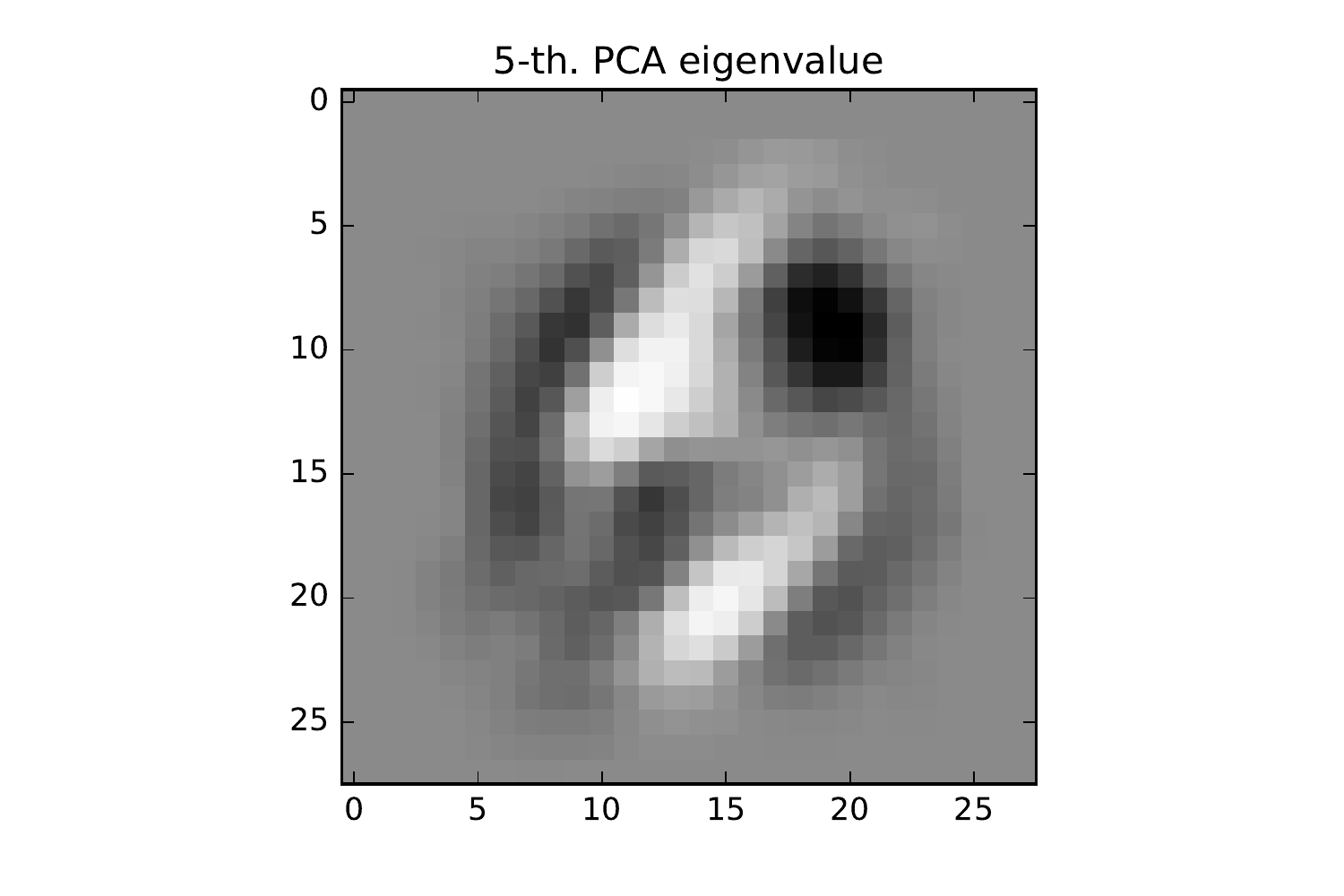}
\includegraphics[width=1.3in]{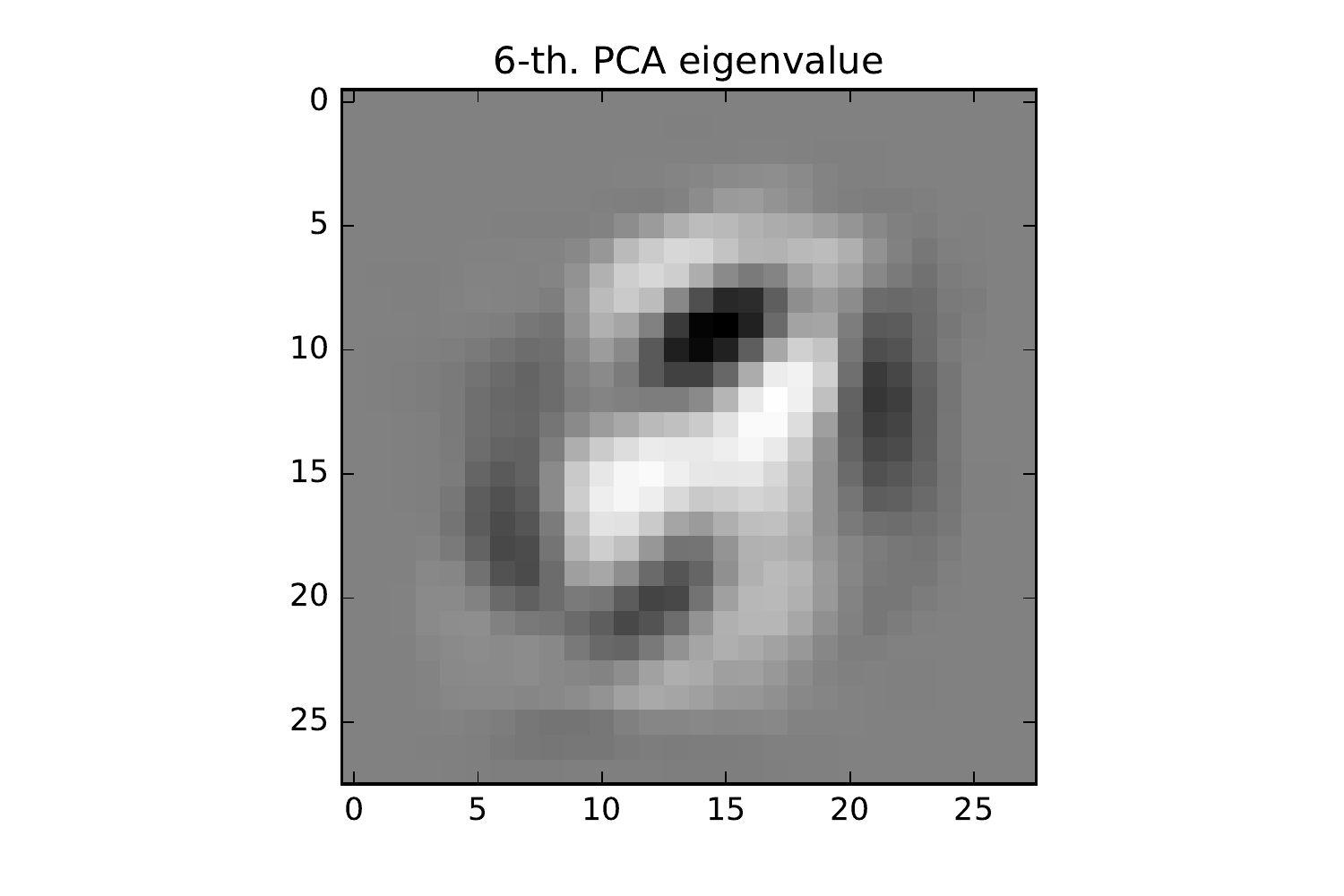}\\
\includegraphics[width=1.3in]{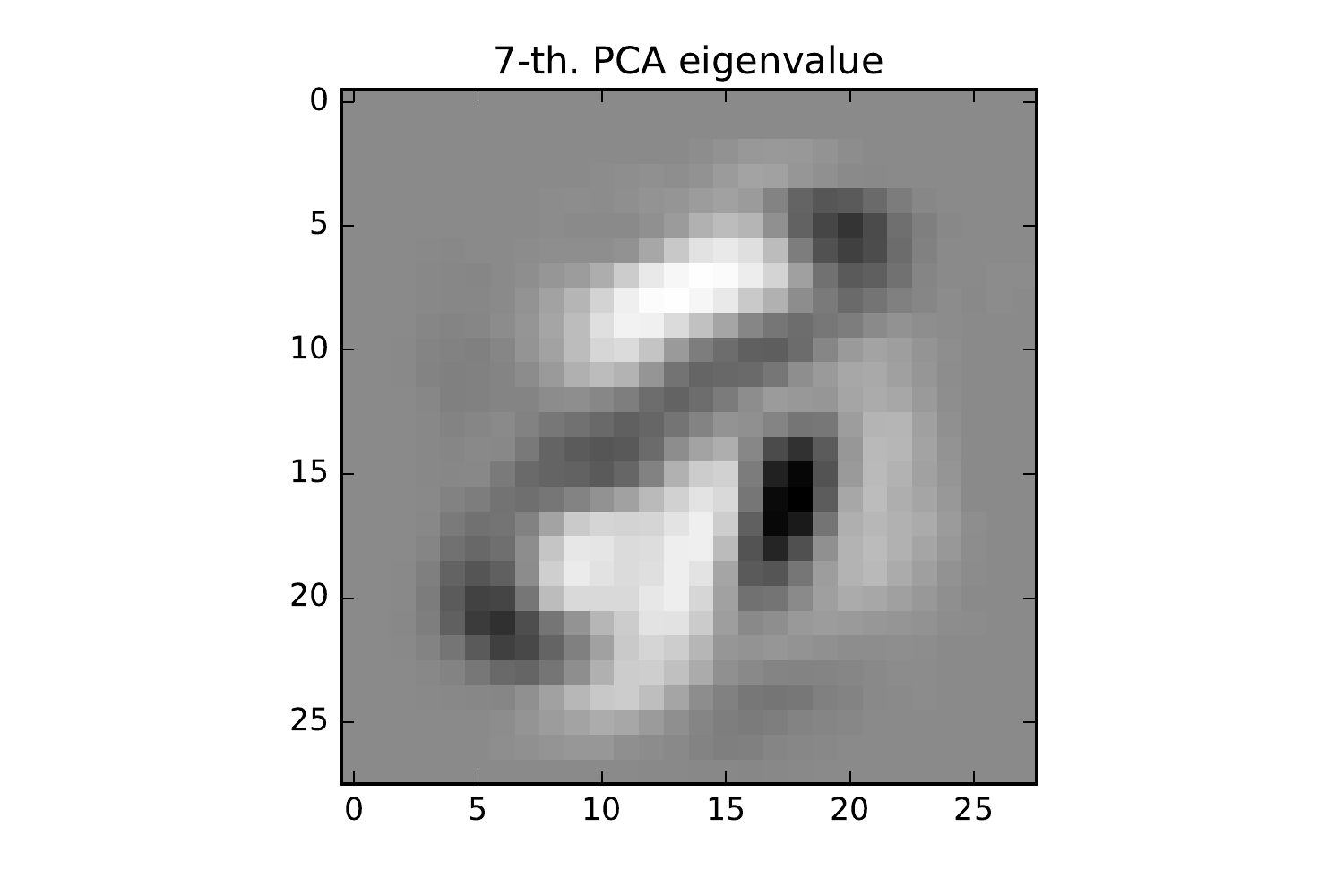}
\includegraphics[width=1.3in]{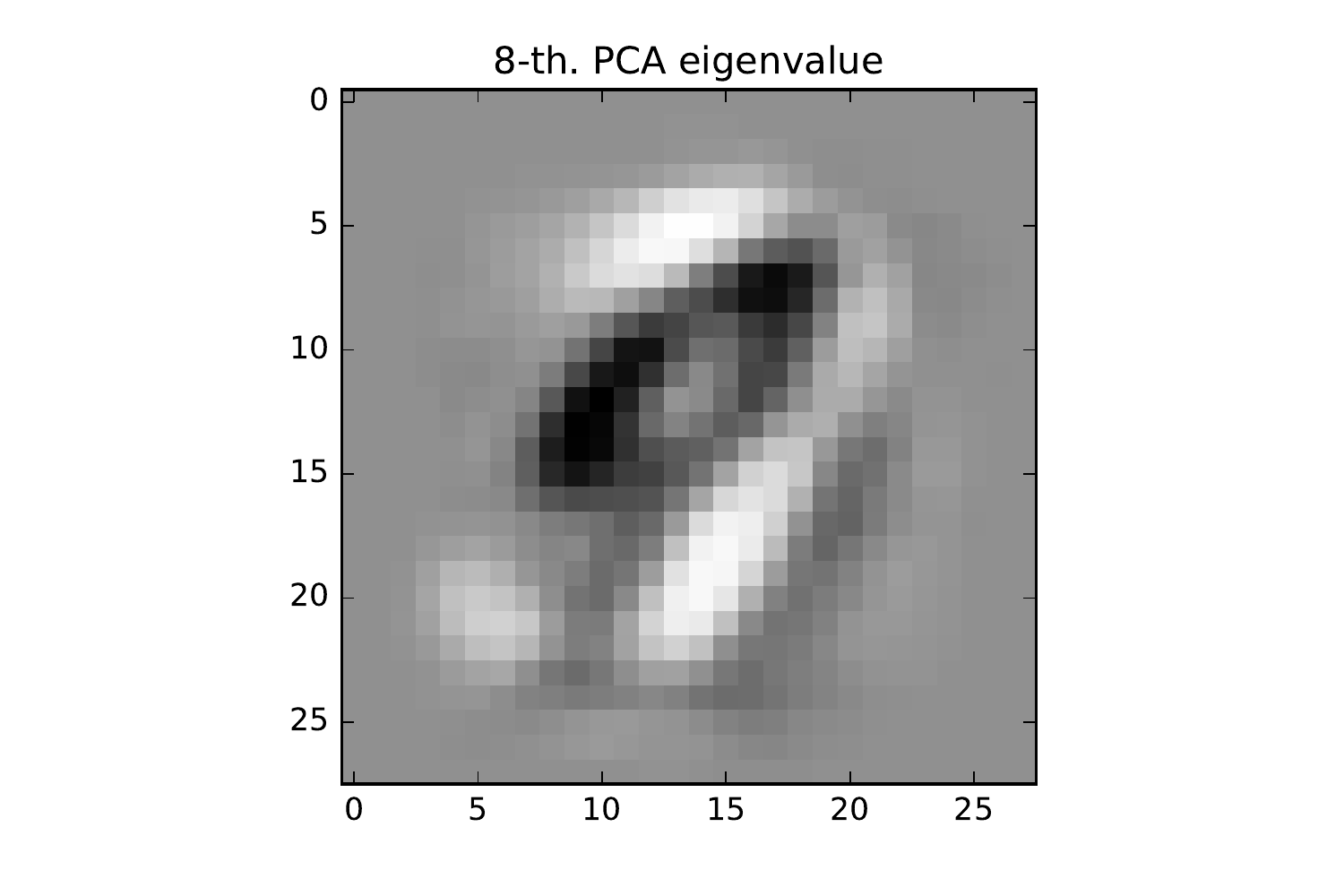}
\includegraphics[width=1.3in]{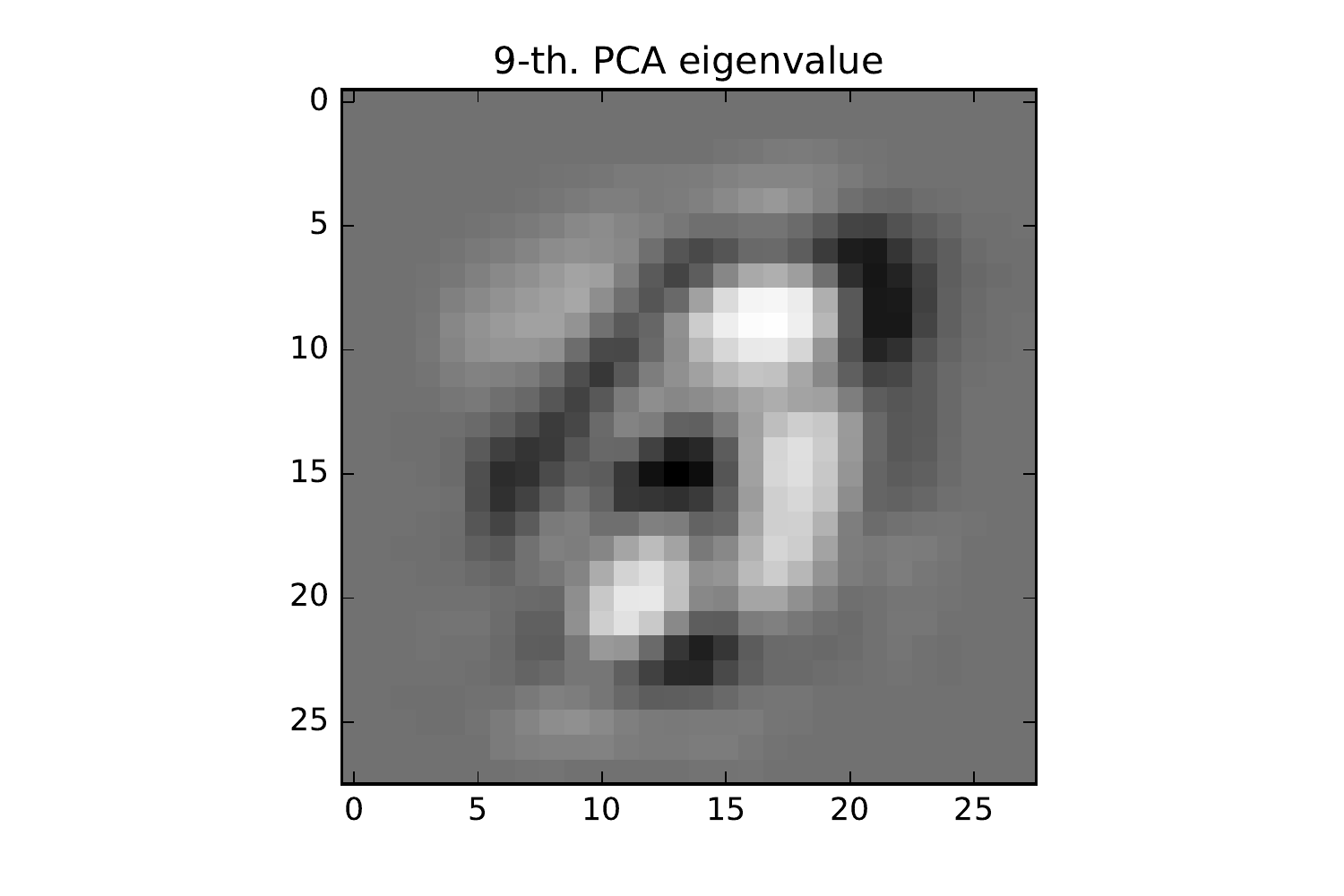}
\caption{\label{fig:faces}First 9 PCA  eigenvectors.}
\includegraphics[width=1.3in]{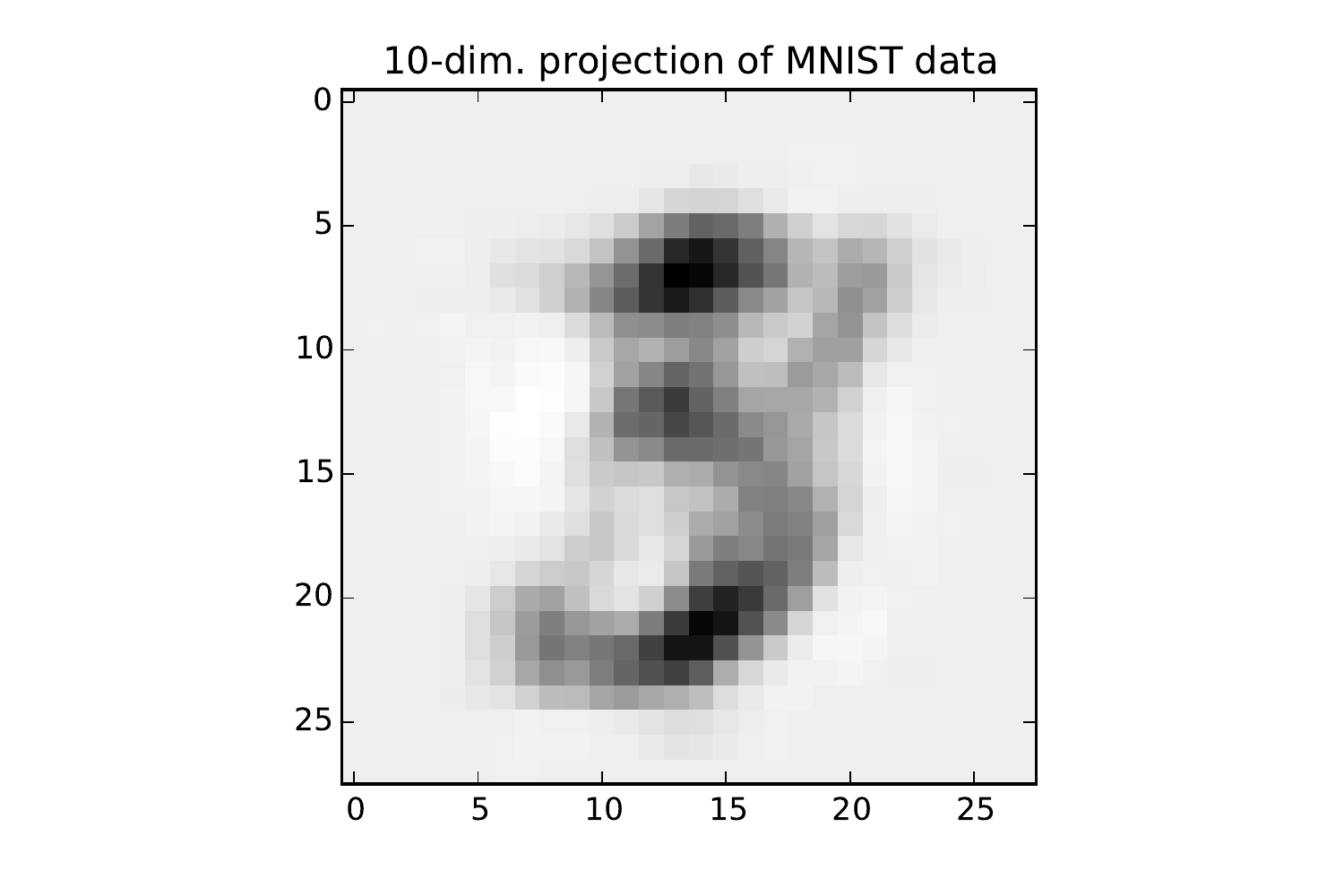}
\includegraphics[width=1.3in]{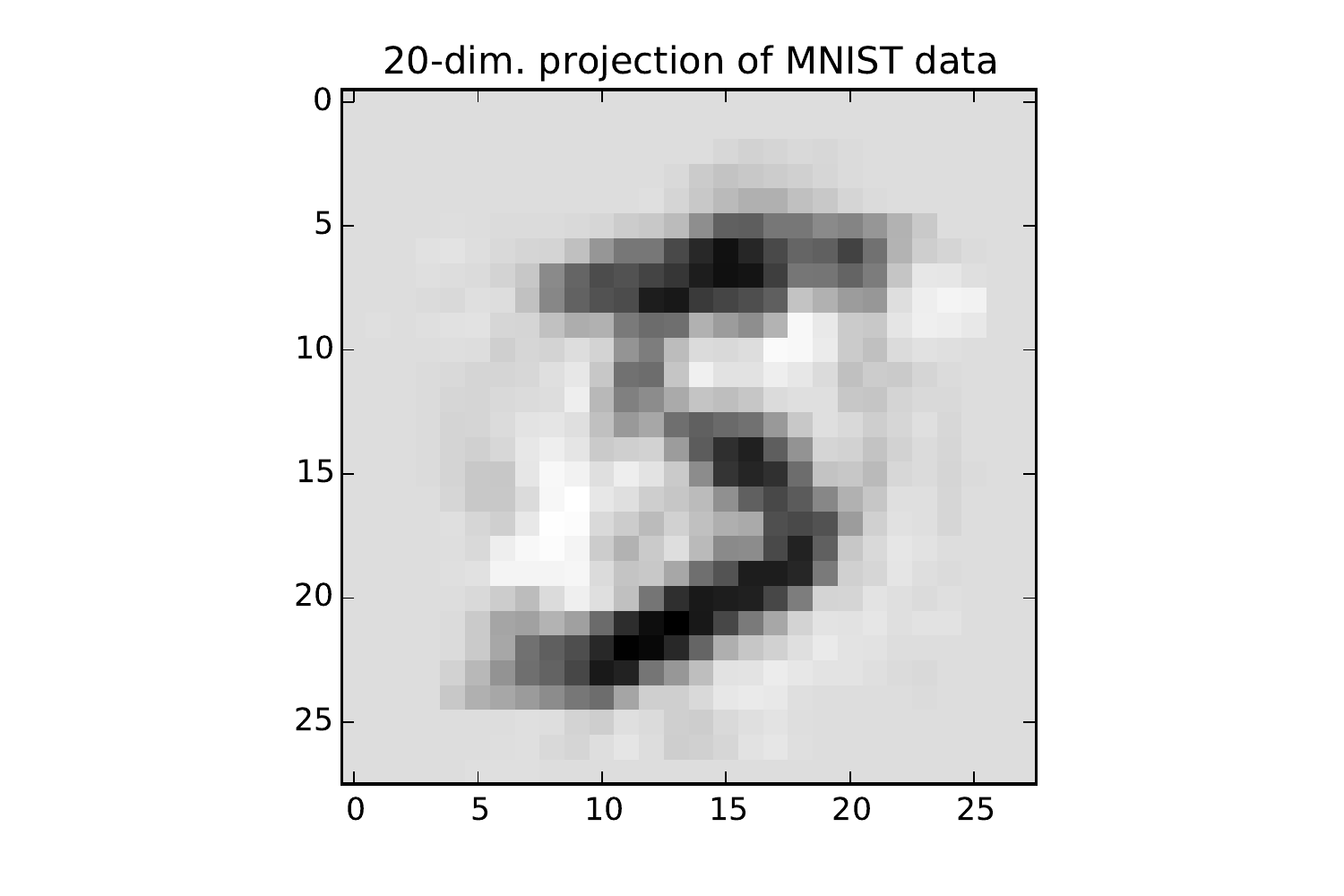}
\includegraphics[width=1.3in]{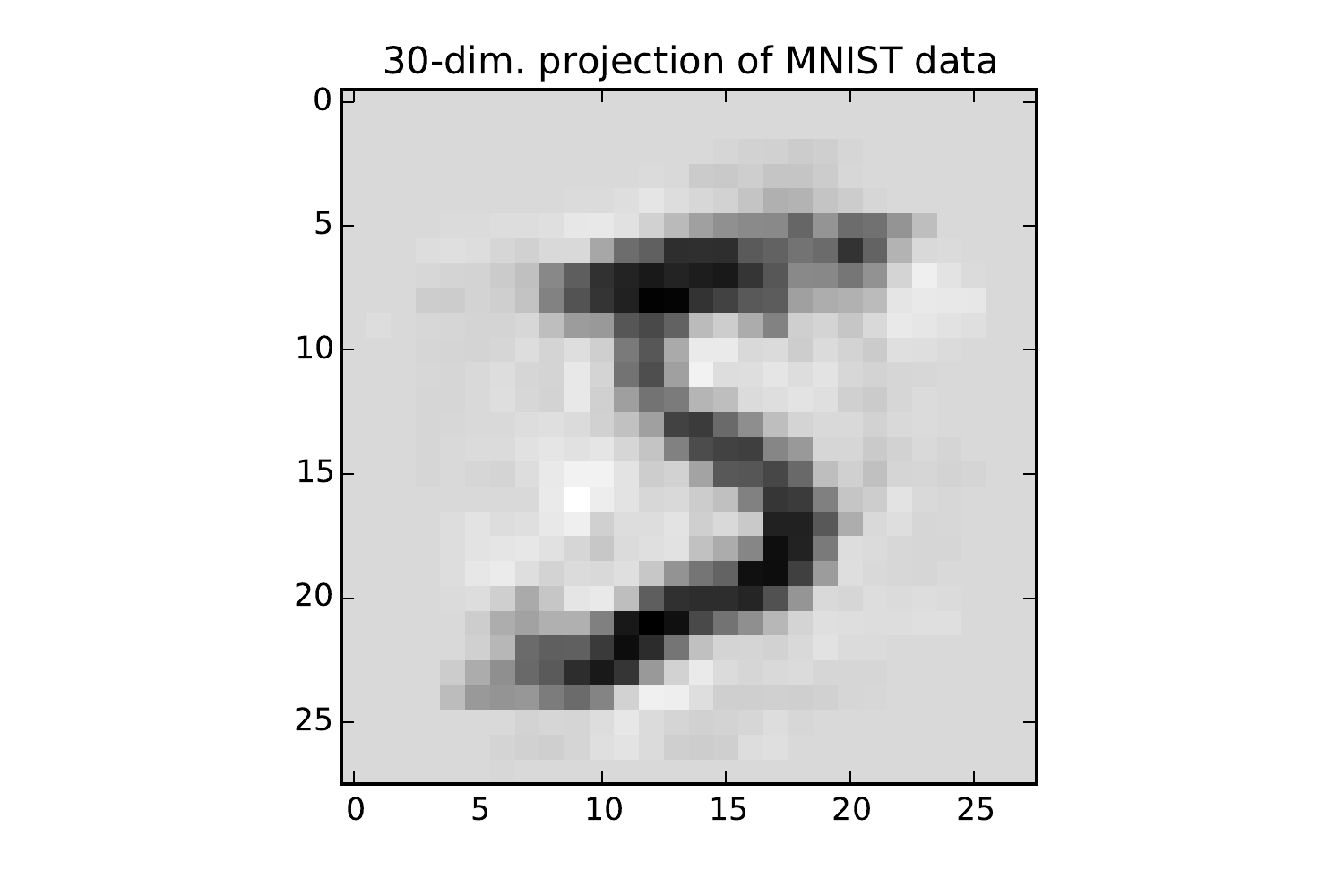}\\
\includegraphics[width=1.3in]{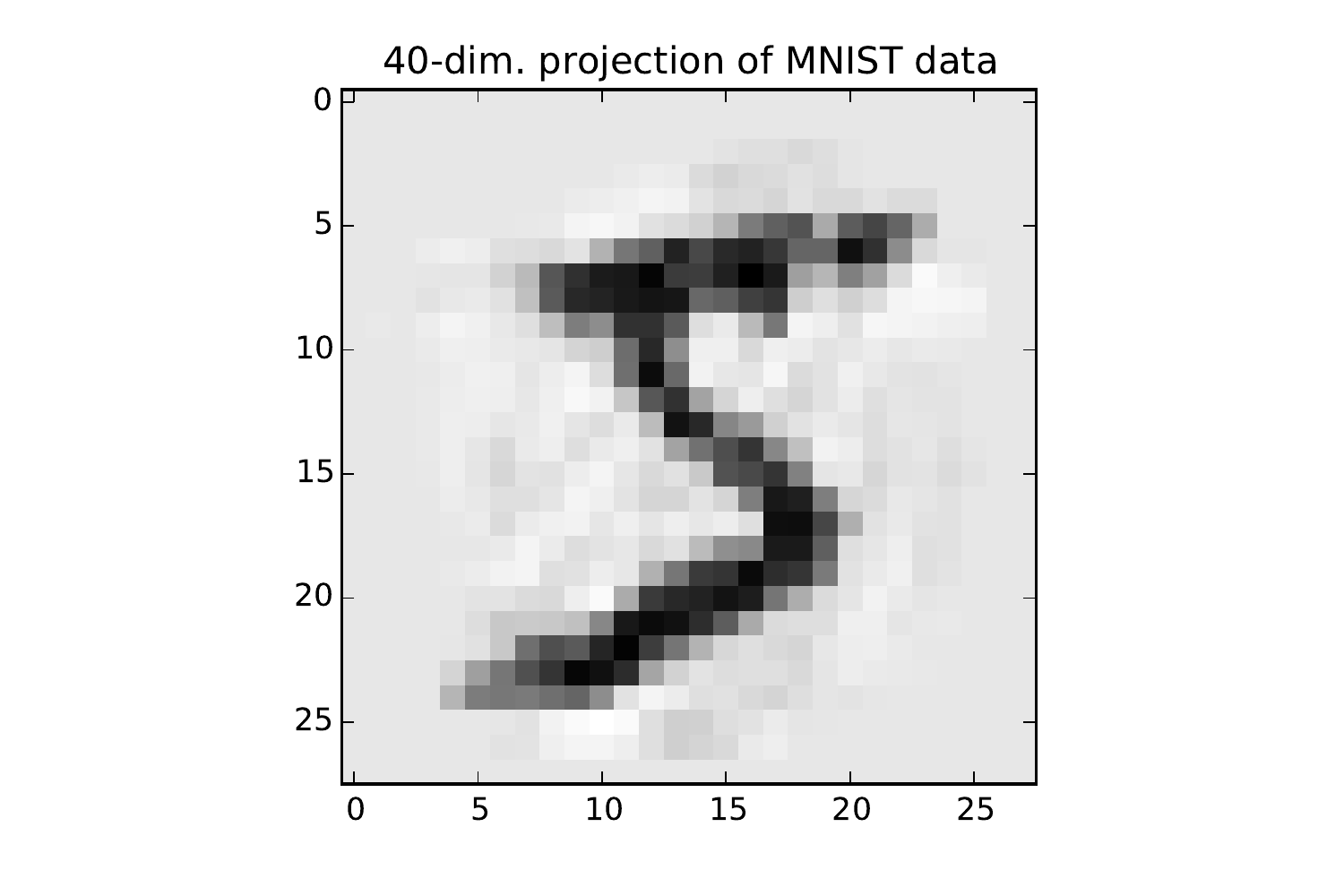}
\includegraphics[width=1.3in]{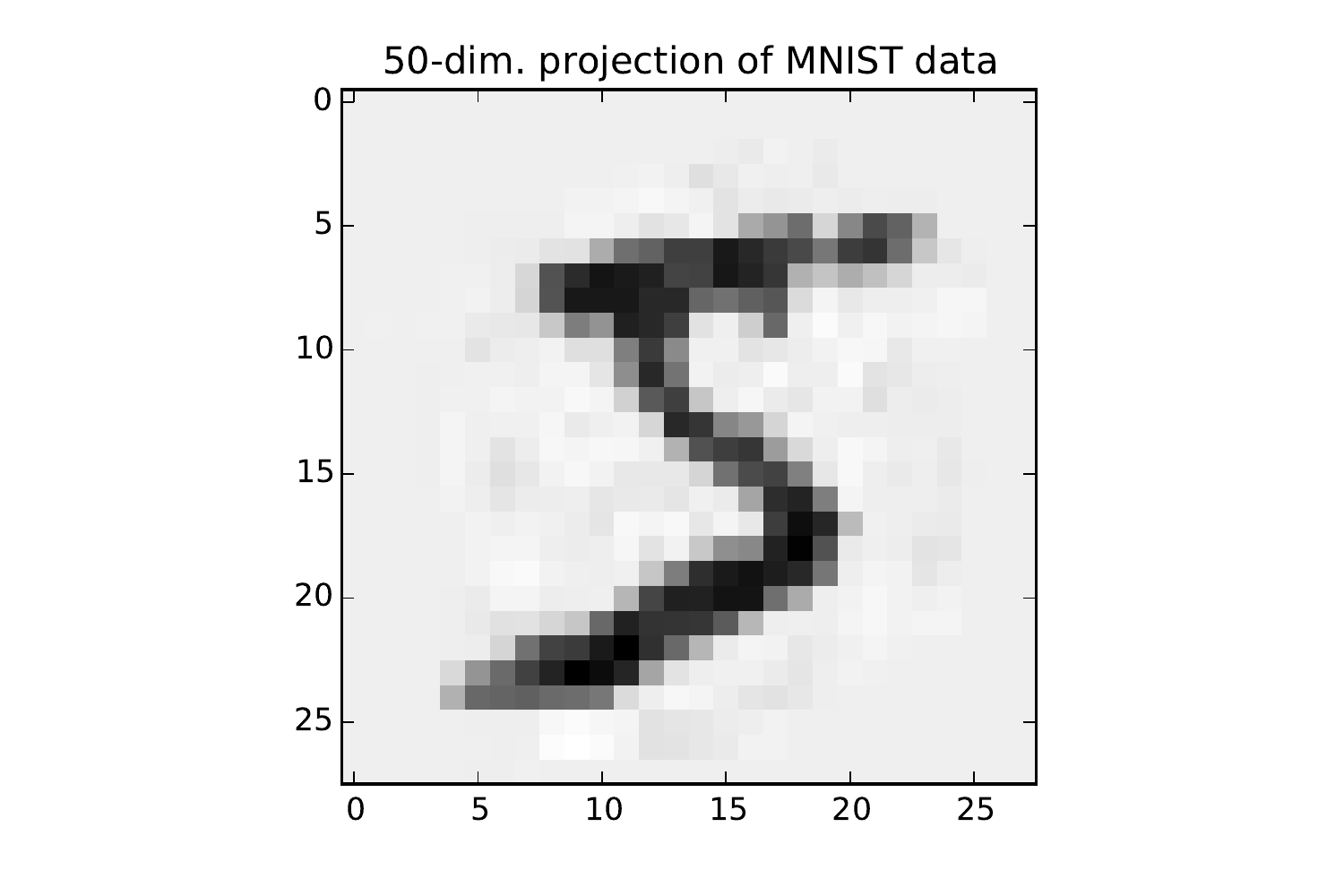}
\includegraphics[width=1.3in]{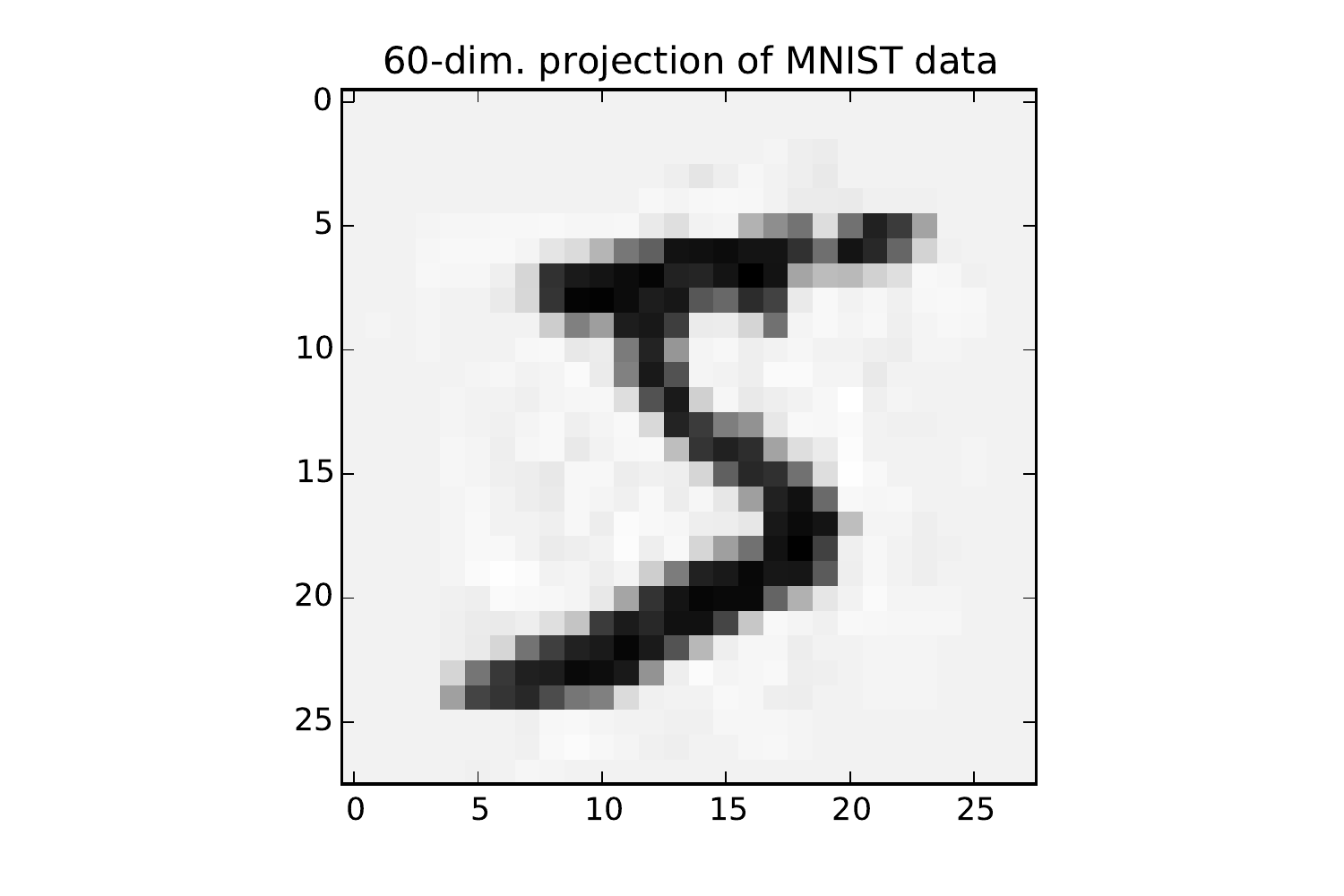}\\
\includegraphics[width=1.3in]{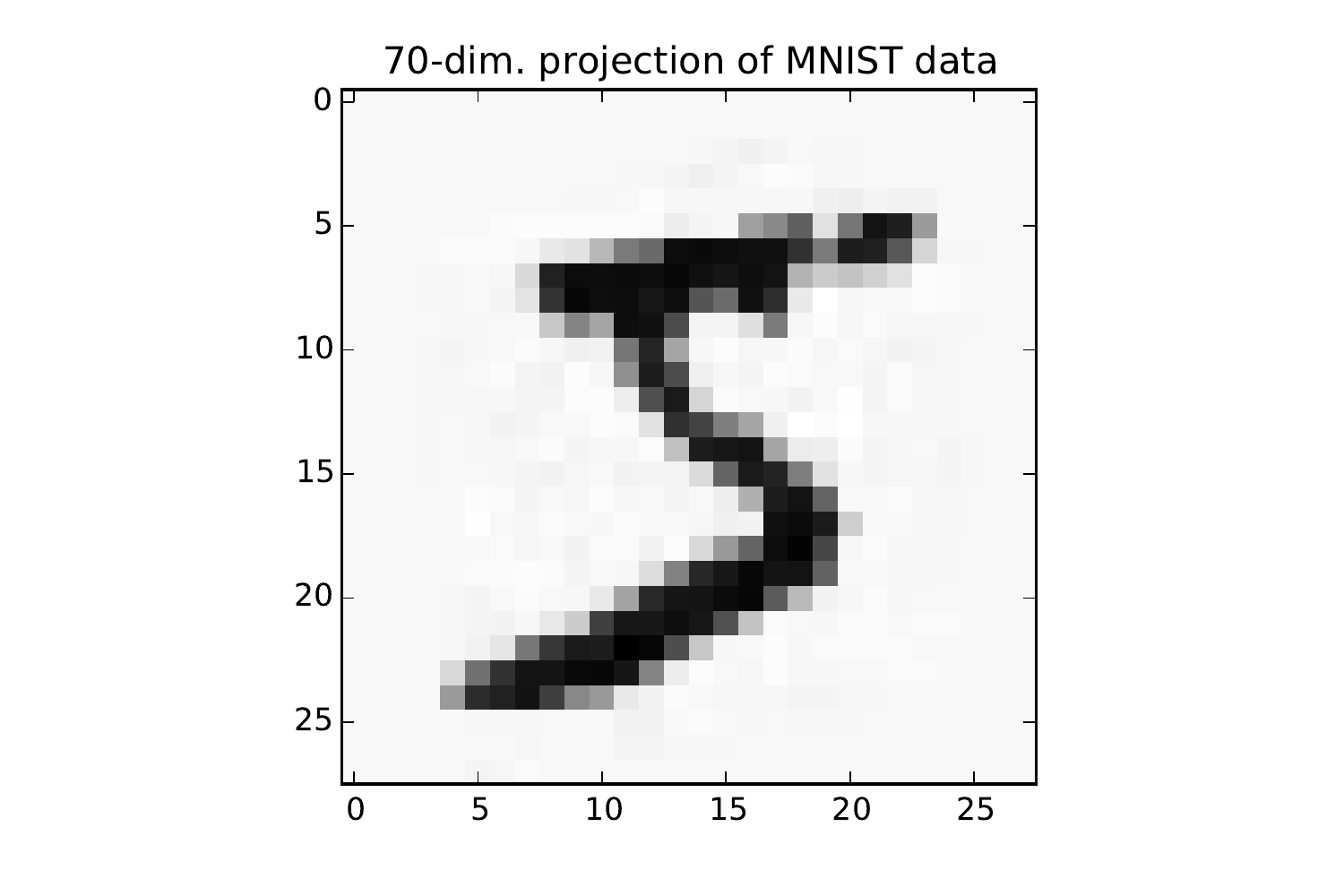}
\includegraphics[width=1.3in]{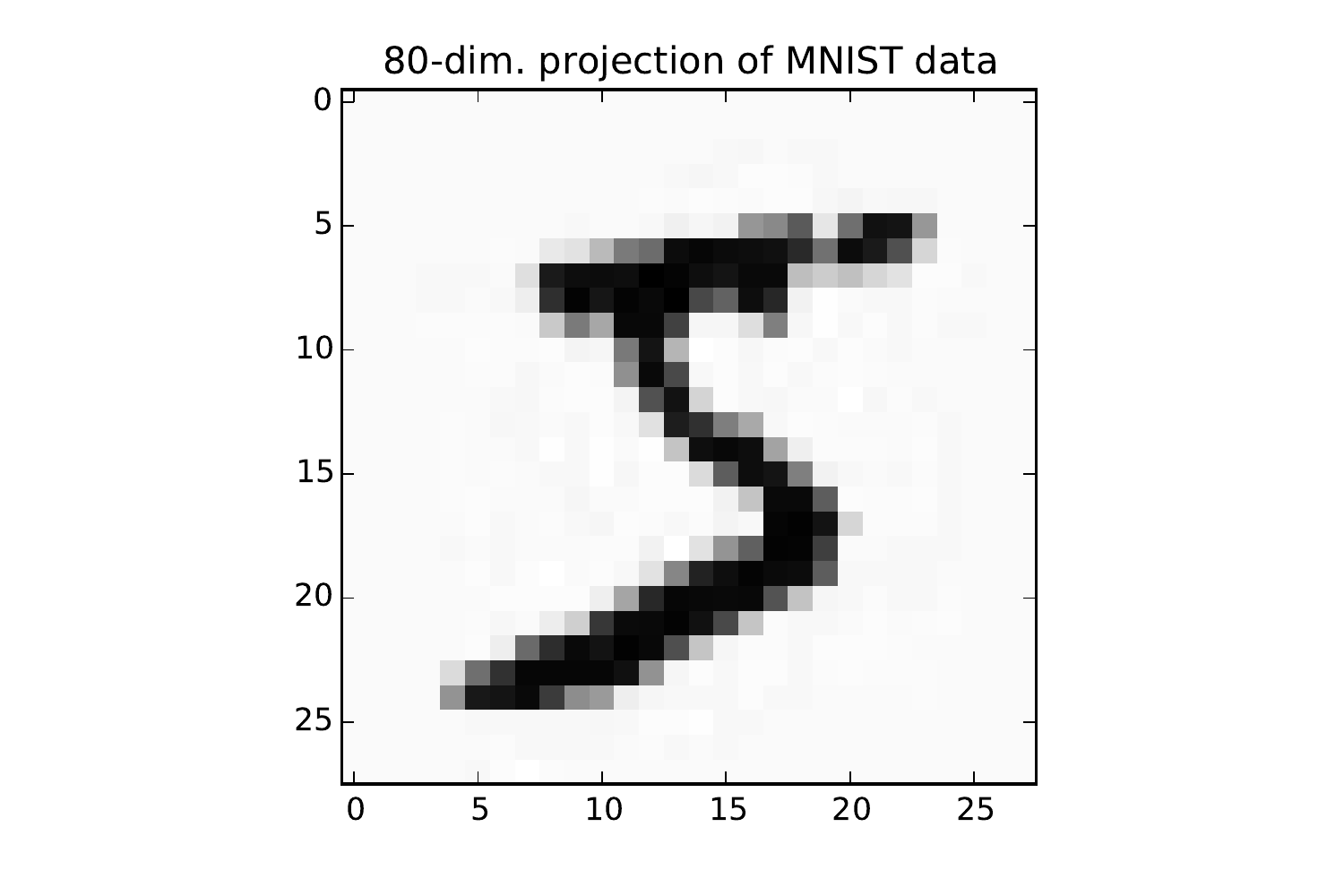}
\includegraphics[width=1.3in]{original.pdf}
\caption{\label{fig:proj}PCA  projections in subspaces of dimensions 10, 20, ... 80 compared to the original 784-dimensional data  in the lower right corner.}
\end{center}
\end{figure}

The PCA method has been used successfully for more than a century. It consists in
identifying directions with largest variance (most relevant directions). It  may allow a drastic reduction of the information necessary to calculate observables. We call $v_i^{(n)}$ the grayscale value of the $i$-th pixel in the $n$-th MNIST sample.
We first define $\bar{v_i}$  as the average grayscale value of the $i$-th pixel over the learning set.
This average is shown in Fig. \ref{fig:ave}.
We can now define the covariance matrix:
\beq
C_{ij}=(1/N)\sum_{n=1}^N(v_i^{(n)}-\bar{v_i})(v_j^{(n)}-\bar{v_j}).
\enq
We can project the original data onto a small dimensional subspace corresponding to the largest eigenvalues of $C_{ij}$.
The first nine
eigenvectors are displayed in Fig. \ref{fig:faces}. The projections in subspaces of dimension 10, 20, ... 80 are shown in Fig. \ref{fig:proj}.

\begin{figure}[hh]
\vskip-40pt
\end{figure}

\section{
RG inspired approximations}
\label{sec:rg}
\begin{figure}[hh]
\begin{center}
\includegraphics[width=0.4\linewidth]{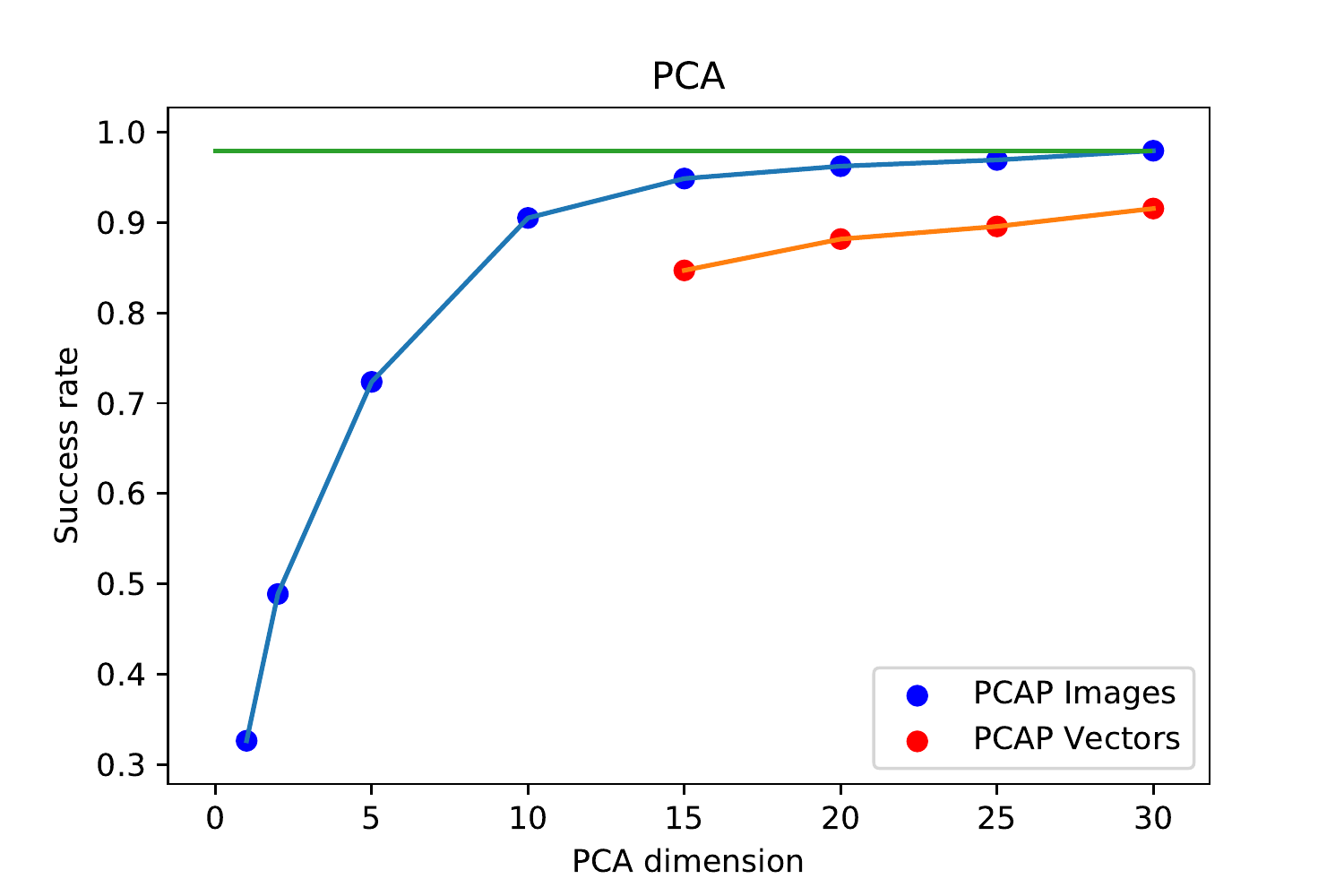}
\end{center}
\label{fig:pcap}
\caption{Success rate for PCA projections of the $28\times28$ images (blue, using the PCA eigenvectors=images)
or using the PCA coordinates only (red, no images) as a function of the dimension of the subspace. The green line represents the
asymptotic value (98 percent) for the original version (dimension 784).}
\end{figure}

We have reconstructed PCA projected training and learning sets either as images (we keep 784 pixels after the projection, as shown in Fig. \ref{fig:proj}, and proceed with the projected images following the usual procedure),
or as abstract vectors (the coordinates of the image in the truncated  eigenvector base without using the eigenvectors, a much smaller dataset). The success rate of these PCA Projections are shown on Fig. \ref{fig:pcap}.

We have used the one hidden layer perceptron with blocked images. First we replaced squares of 4 four pixels by a single pixel carrying the average value of the four blocked pixels. Using the $14\times14$ blocked pictures with 49 hidden variables, the success rate
goes down slightly (97 percent). Repeating once, we obtain $7\times7$ images. With 25 hidden variables, we get a success rate of 92 percent.

As mentioned before, we can also replace the grayscale pixels by black and white pixels. This barely affects the performance (97.6 percent) but
diminishes the configuration space from $256^{784}$ to $2^{784}$ and allows a Restricted Boltzmann Machine treatment.

We have considered the hierarchical approximation where each hidden variable is only connected to a single $2\times2$ block of visible variables (pixels):
\beq
W^{(1)}_{lj}v_j\rightarrow W^{(1)}_{l,\alpha} v_{l,\alpha}
\enq
with $\alpha$ = 1, 2, 3, 4 are the position in the $2\times 2$ block and $l$ = 1, ...,196 the labeling of the blocks.
Even though the number of parameters that we need to determine with the gradient method is significantly smaller (by a factor 196), the performance remains 0.92.
 A generalization with 4x4 blocks leads to a 0.90 performance with 1/4 as many weights.
 This simplified version can be used as a starting point for a full gradient search (pretraining), but the hierarchical structure (sparcity of $W_{ij}$) is robust and remains visible during the training. This pretraining breaks the huge permutation symmetry of the hidden variables.

\section{Transition to the 2D Ising model}
\label{sec:ising}

The MNIST data has a typical size built in  the images, namely the width of the lines and UV details can be erased without drastic effects  until that size is reached.
One can think that the various digits are separate ``phases", but there is nothing like a critical point connected to all the phases.
It might be possible to think of the average $\bar{v}_i$ as a fixed point. However, there are no images close to it.
In order to get images that can be understood as close to a critical point, we will consider the images of  worm configurations for the 2D Ising model at different $\beta=1/T$, some close to the critical value.
The graphs in this section have been made by Sam Foreman. 

The worm algorithm \cite{PhysRevLett.87.160601} allows us to sample the high temperature contributions, whose statistics are governed by the number of active bonds in a given configuration. An example of an equilibrium configuration is shown in Fig. \ref{fig:worm}. 
We implemented a `coarse-graining` procedure where the lattice is divided into blocks of $2\times 2$ squares, essentially reducing the size of each linear dimension by two. Each $2\times 2$ square is then `blocked` into a single site, where the new external bonds in a given direction are determined by the number of active bonds exiting a given square.
If a given block has one external bond in a given direction, the
blocked site retains this bond in the blocked configuration,
otherwise it is ignored. This is illustrated on the right side of Fig. \ref{fig:worm}.
\begin{figure}[h!]
    \begin{center}
        \includegraphics[width=2.in]{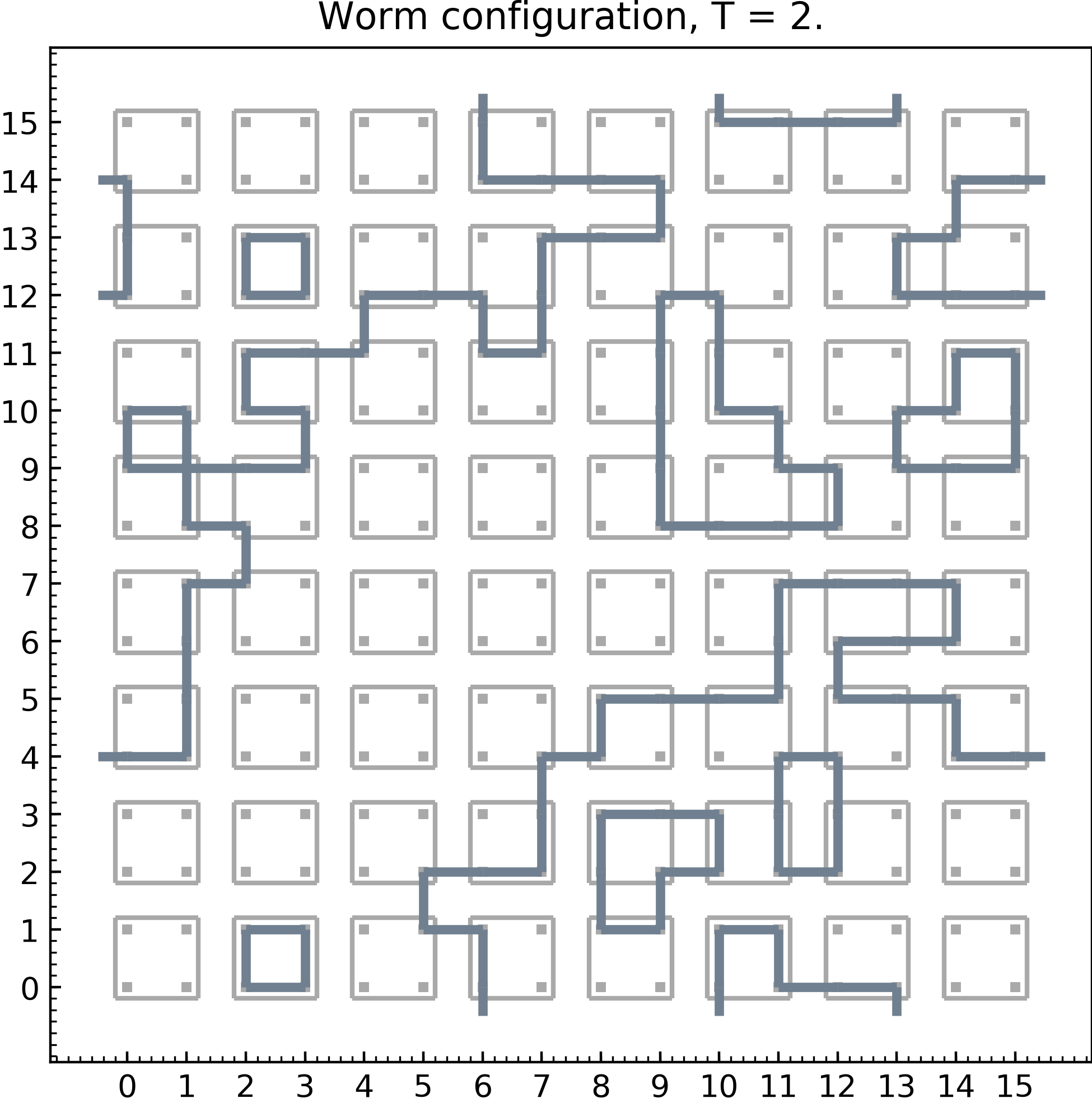}%
        \quad
        \includegraphics[width=2.in]{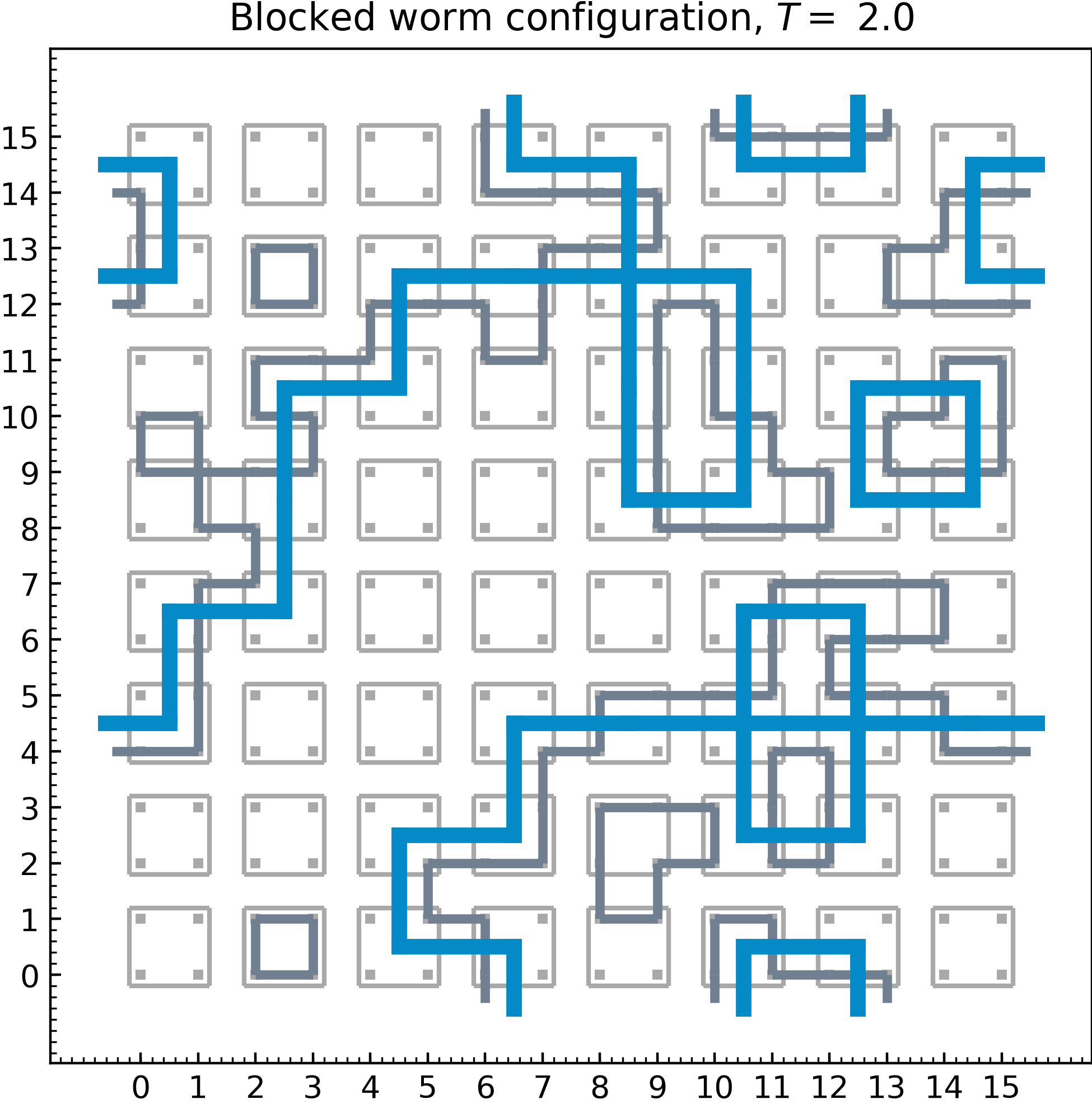}%
        \label{fig:worm}
    \end{center}
    \caption{\label{fig:worm}Example of legal high temperature contribution also called worm (left). All the paths close due to periodic boundary conditions.
Example of ``worm blocking`` applied to the same configuration  (right).}
\end{figure}
 
\begin{figure}[!h]
\label{fig:cv_new}

\centering
    \subfigure{
    \centering
      \includegraphics[width=0.4\textwidth]{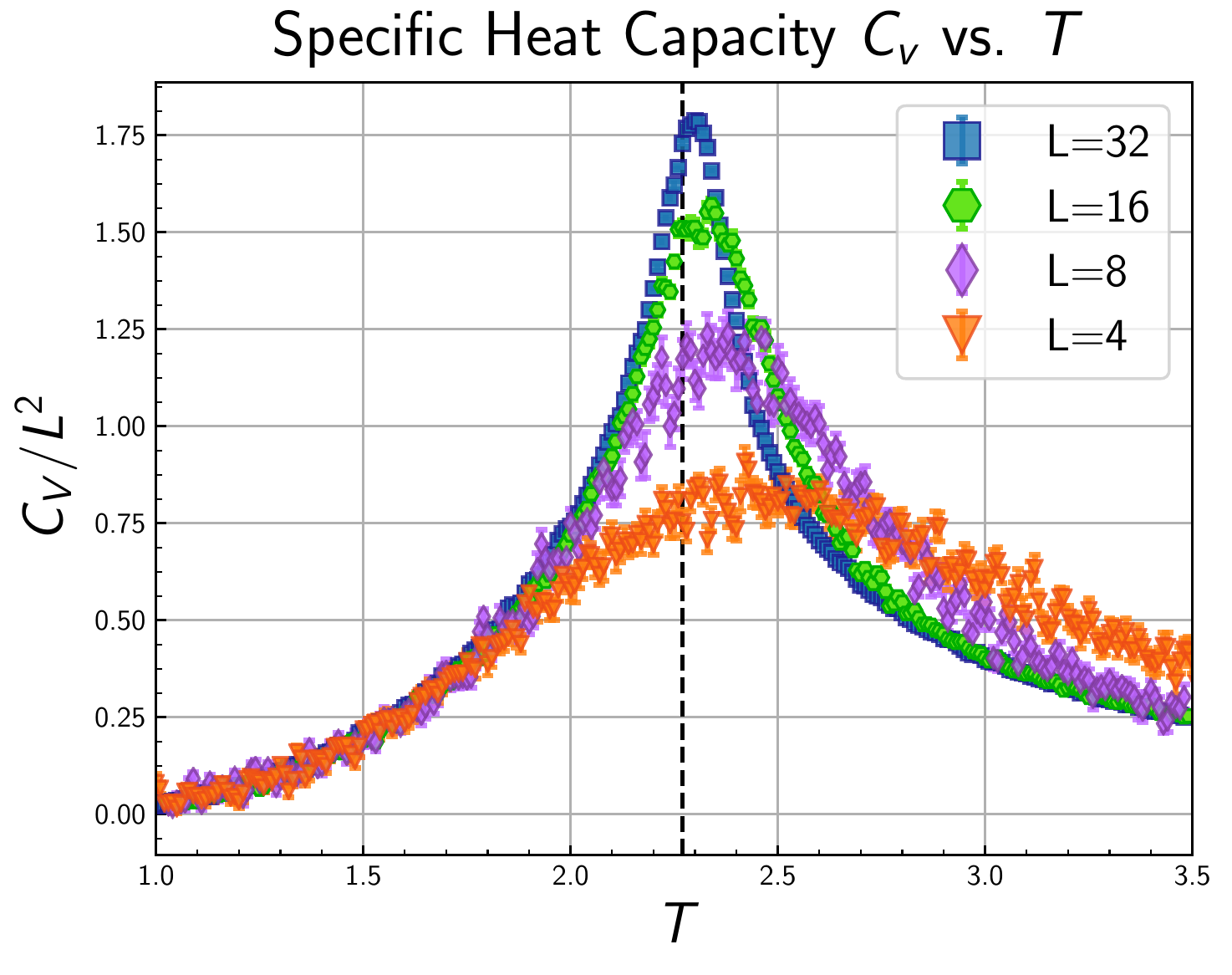}
    }
    \subfigure{
    \centering
      \includegraphics[width=0.4\textwidth]{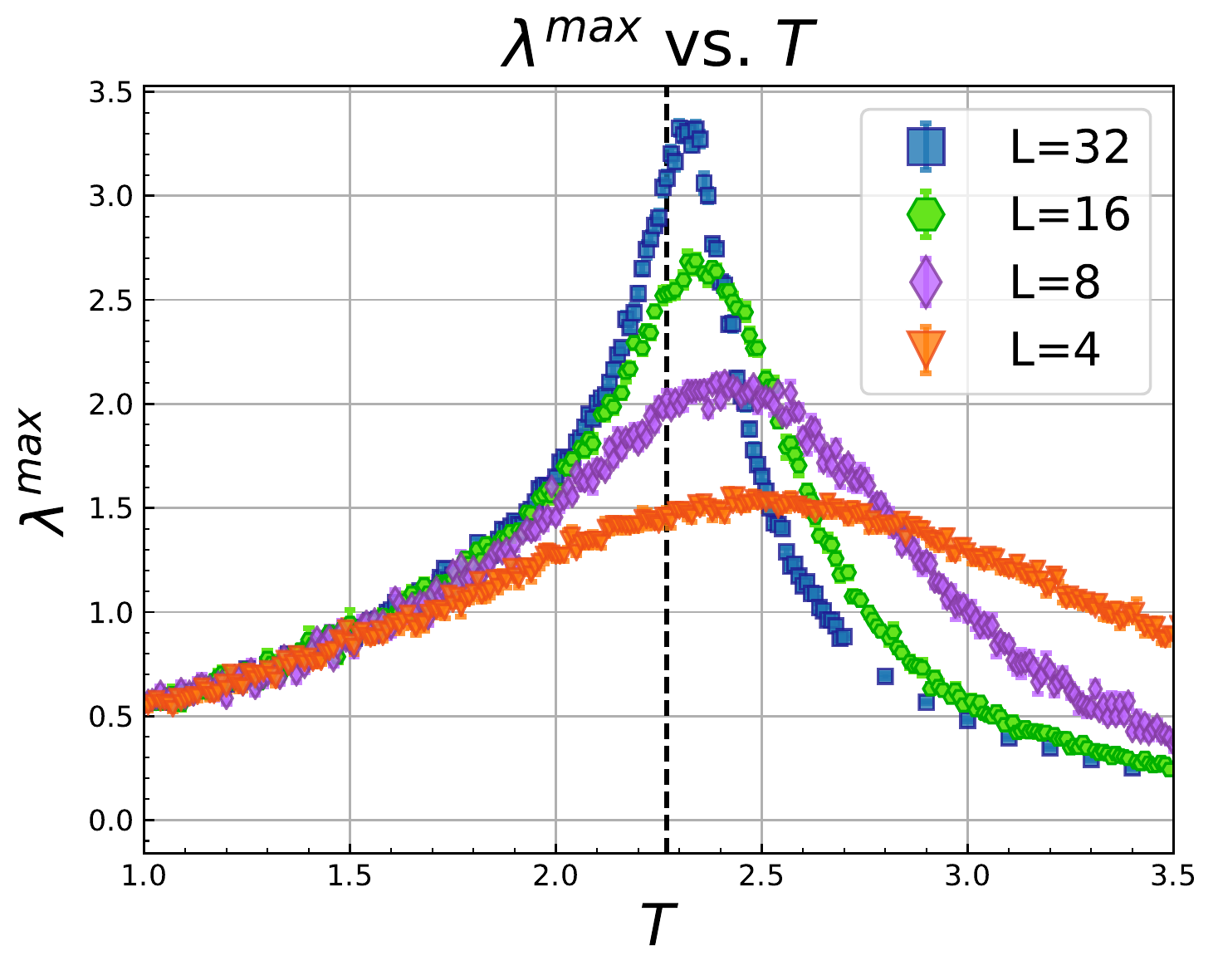}
    }
    \caption{ Specific heat capacity (left) and largest PCA eigenvalue (right) for various lattice sizes using improved statistics compared to the figures presented at the conference.}
\end{figure}

 The worm algorithm allows statistically exact calculations of the specific heat.
In order to observe finite size effects, we performed this analysis for lattice sizes $L = 4, 8, 16, 32$. Using arguments that will be presented elsewhere \cite{inprogress}, we conjectured that near criticality, the largest PCA eigenvalue $\lambda_{max}$  is proportional to the specific heat per unit of volume, with a proportionality constant $\frac{2}{3} \left(\ln(1+\sqrt{2})\right)^2\approx 0.52$. The good agreement is illustrated in Fig. \ref{fig:cv_new}. 

To cross check the accuracy of the worm for some observables, the TRG was used.  We calculated $\langle N_{b} \rangle$ for a blocked and unblocked lattice using TRG, and calculated $\langle (N_{b} - \langle N_{b} \rangle )^{2} \rangle$ for the unblocked case, and good agreement was found between the worm and the TRG.  The TRG also provides a clear picture of what bond numbers are associated with which states as opposed to editing pictures by setting pixels to chosen values, such as what happens in the blocked case for a block of four sites.  This correspondence  makes the picture of RG for the configurations sturdier.

Results for the leading PCA eigenvalue and specific heat for blocked configurations were found qualitatively similar to the unblocked results. The blocking procedure is approximate. Systematically improvable approximations can be constructed with the TRG method \cite{prb87,efratirmp,prd88}. Improvement of the blocking method used here were developed after the conference and will be presented elsewhere \cite{inprogress}.

\section{Conclusions}
ML has a clear  RG flavor. A general question in this field is how to extract relevant features from noisy pictures (``configurations").
The RG ideas allow to reduce the complexity of the perceptron algorithms for the MNIST data.
Direct use of Tensor RG methods are being used for the ML treatment of worm configurations of the 2D Ising model near criticality
and will be presented elsewhere \cite{inprogress}. Sam Foreman has been developing methods to identify the side of transition. He has improved 
a fully-connected perceptron approach by utilizing a convolutional neural network (ConvNet) for supervised learning. ConvNets have wide applications in image recognition problems, and in particular, are known to perform extremely well at classification tasks \cite{conv_net}. See also Ref. \cite{tanaka}. 
The methods could be extended to identify the temperature. For large enough lattices, this procedure could borrow from the master field ideas presented at the conference \cite{masterfield}.
Judah Unmuth-Yockey has made progress with restricted Boltzmann machines (RBMs) learning on configurations of Ising spins.  Here it was found that the RBMs are capable of learning the local two-body interaction of the original model.  The Hamiltonian of the learned RBM is then comprised of effective $n$-body interactions. 
Sam Foreman  improve upon our initial approach (
ML could benefit Lattice Field Theory: how to compress configurations, find features or updates \cite{update}.

This work was supported by the U.S. Department of Energy (DOE), Office of Science, Office of High Energy Physics, under Award Numbers DE- SC0013496 (JG) and DE-SC0010113 (YM).


\end{document}